\documentclass[11pt]{article}
\usepackage[utf8]{inputenc}
\usepackage{graphicx}
\usepackage{amsmath}
\usepackage{amssymb}
\usepackage{tabularx}
\usepackage[dvipsnames]{xcolor}
\usepackage{mwe}
\usepackage{wrapfig}
\usepackage{multirow}
\usepackage{float}
\usepackage[hyphens]{url}
\usepackage{hyperref}

\RequirePackage[left=1in,%
                right=1in,%
                top=1in,%
                bottom=1in,%
                headheight=12pt,%
                letterpaper]{geometry}%
\usepackage{enumitem}
\usepackage{booktabs}
\usepackage[most]{tcolorbox}
\usepackage{caption}
\usepackage{url}
\usepackage{mathtools}

\DeclareMathOperator{\sclip}{sclip}
\DeclareMathOperator{\smin}{smin}
\DeclareMathOperator{\smax}{smax}
\newcommand{\ssign}{\operatorname{sign}_\varepsilon}
\newtheorem{remark}{Remark}

\title{%
    \Large\bfseries
Dynamic Modeling of Transmission Networks, Grid-Following Inverters, and UPS-Based AI Data Centers
}

\author{%
\large
Muhammad Hamza Ali and Amritanshu Pandey\\[0.5ex]
\normalsize
Department of Electrical and Biomedical Engineering\\
University of Vermont\\
Burlington, Vermont
}

\date{}

\begin{document}

\maketitle

\begin{tcolorbox}[colback=gray!10, colframe=gray!45, boxrule=0.5pt, arc=2pt, left=8pt, right=8pt, top=6pt, bottom=6pt]
\small
\textbf{Supplementary material.} This document is the supplementary material of the paper
\textit{``Fast Adaptive Multirate Waveform Relaxation for Multi-Timescale Power System Simulations''} by
M.~H.~Ali and A.~Pandey (submitted for publication). It collects the equations and parameters of every model used in
that paper, namely the transmission network with synchronous generators, the grid-following inverters, and the
double-conversion UPS-based AI data centers, so that its results can be reproduced. Section, equation, and figure
numbers refer to this document.
\end{tcolorbox}

This appendix details the dynamic and algebraic models used to construct the slow and fast subsystems in the paper \textbf{Section~II-A}. We formulate the slow subsystem using the transmission network and synchronous generators, and model the fast subsystem using average models of grid-following inverters and double-conversion UPS-based AI data centers. The two subsystems interact through the voltage magnitude, phase angle, active power, and reactive power at each coupling bus. The following sections present the governing equations and
state and algebraic variables of each subsystem.

\section{Nomenclature}\label{sec:nomenclature}
\subsection*{Index Sets}
\begin{tabularx}{\linewidth}{@{}lX@{}}
\toprule
Symbol & Description \\
\midrule
$i\in\mathcal{N}_g=\{1,\dots,N_g\}$ & synchronous generators \\
$i,k\in\mathcal{N}_b=\{1,\dots,N_b\}$ & network buses; $j$ denotes the bus hosting a device \\
$i\in\mathcal{N}_l=\{1,\dots,N_l\}$ & network loads \\
$i\in\mathcal{N}_{\mathrm{inv}}=\{1,\dots,N_{\mathrm{inv}}\}$ & grid-following inverters \\
$i\in\mathcal{N}_{\mathrm{dc}}=\{1,\dots,N_{\mathrm{dc}}\}$ & data centers \\
$i\in\mathcal{N}_{\mathrm{rect}}^{\mathrm{dc}}$ & data-center UPS rectifiers \\
$i\in\mathcal{N}_{\mathrm{bat}}^{\mathrm{dc}}$ & data-center UPS batteries \\
$i\in\mathcal{N}_{\mathrm{conv}}^{\mathrm{dc}}$ & data-center bidirectional DC--DC converters \\
\bottomrule
\end{tabularx}

\subsection*{Transmission Network and Synchronous Generator}
\begin{tabularx}{\linewidth}{@{}lX@{}}
\toprule
Symbol & Description \\
\midrule
$V_{j}^{b},\ \delta_{j}^{b}$ & voltage magnitude and phase angle of bus $j$ \\
$\delta_{ik}^{b}$ & voltage angle difference $\delta_i^b-\delta_k^b$ \\
$G_{ik},\ B_{ik}$ & conductance and susceptance between buses $i$ and $k$ \\
$\delta_i^{g}$ & internal rotor angle of generator $i$ \\
$\Delta\omega_i^{g}$ & per-unit rotor-speed deviation of generator $i$ \\
$\omega_0=2\pi f_0$ & nominal angular frequency \\
$E_{i}'$ & constant internal voltage of generator $i$ \\
$X_{i}^{d'}$ & transient reactance of generator $i$ \\
$H_i^{g},\ D_i^{g}$ & inertia constant and damping coefficient of generator $i$ \\
$P_{i}^{m}$ & mechanical input power; $\underline{P}_{i}^{m},\overline{P}_{i}^{m}$ its bounds \\
$P_{i}^{e},\ Q_{i}^{e}$ & electrical active and reactive power of generator $i$ \\
$P_{i}^{g},\ Q_{i}^{g}$ & generator active/reactive power aggregated at bus $i$ \\
$P_{i}^{d},\ Q_{i}^{d}$ & active and reactive load demand at bus $i$ \\
$T_{i}^{\mathrm{gov}}$ & governor-turbine time constant \\
$P_{i}^{\mathrm{ref}}$ & governor power setpoint \\
$R_i$ & governor droop \\
\bottomrule
\end{tabularx}

\subsection*{Grid-Following Inverter}
\begin{tabularx}{\linewidth}{@{}lX@{}}
\toprule
Symbol & Description \\
\midrule
$v_{i,g}^{d},\ v_{i,g}^{q}$ & grid-side terminal voltage in the $dq$ frame \\
$i_{i,g}^{d},\ i_{i,g}^{q}$ & grid-side current in the $dq$ frame \\
$v_{i,g}^{q,\mathrm{fil}}$ & low-pass-filtered quadrature-axis voltage \\
$i_{i,g}^{d,\mathrm{fil}},\ i_{i,g}^{q,\mathrm{fil}}$ & filtered $dq$ currents \\
$i_{i,g,\mathrm{ref}}^{d},\ i_{i,g,\mathrm{ref}}^{q}$ & $dq$ current references \\
$\theta_{i,\mathrm{PLL}}$ & PLL-estimated grid angle \\
$I_{i,\mathrm{PLL}}$ & PLL integrator state \\
$\Delta_{i,\mathrm{PLL}}$ & PLL PI output (per-unit frequency deviation) \\
$\hat{\omega}_{i}=\dot{\theta}_{i,\mathrm{PLL}}$ & PLL-estimated angular frequency \\
$P_{i,g},\ Q_{i,g}$ & instantaneous active and reactive power \\
$P_{i,g}^{\mathrm{fil}},\ Q_{i,g}^{\mathrm{fil}}$ & filtered active and reactive power \\
$P_{i,\mathrm{ref}},\ Q_{i,\mathrm{ref}}$ & active and reactive power references \\
$I_{i,P}^{d},\ I_{i,Q}^{q}$ & power-controller integrator states \\
$I_{i,\mathrm{cc}}^{d},\ I_{i,\mathrm{cc}}^{q}$ & current-controller integrator states \\
$\epsilon_{i}^{d},\ \epsilon_{i}^{q}$ & current-controller PI outputs \\
$e_{i,\mathrm{inv}}^{d},\ e_{i,\mathrm{inv}}^{q}$ & inverter terminal voltage commands \\
$R_f^{\mathrm{inv}},\ L_f^{\mathrm{inv}}$ & inverter output-filter resistance and inductance \\
$T_{f,\mathrm{PLL}}^{\mathrm{inv}},\ T_{f,\mathrm{PQ}}^{\mathrm{inv}},\ T_{f,\mathrm{ig}}^{\mathrm{inv}}$ & PLL, power-loop, and current-loop filter time constants \\
$k_{P,\mathrm{PLL}}^{\mathrm{inv}},\ k_{I,\mathrm{PLL}}^{\mathrm{inv}}$ & PLL PI gains \\
$k_{P,\mathrm{PQ}}^{\mathrm{inv}},\ k_{I,\mathrm{PQ}}^{\mathrm{inv}}$ & power-controller PI gains \\
$k_{P,\mathrm{ig}}^{\mathrm{inv}},\ k_{I,\mathrm{ig}}^{\mathrm{inv}}$ & current-controller PI gains \\
\bottomrule
\end{tabularx}

\subsection*{Data-Center Transformer}
\begin{tabularx}{\linewidth}{@{}lX@{}}
\toprule
Symbol & Description \\
\midrule
$R_{i,k},\ X_{i,k}$ & leakage resistance and reactance of transformer $k\in\{1,2\}$ (T1: HV/MV, T2: MV/LV) \\
$R_{i,T},\ X_{i,T}$ & lumped series resistance and reactance of the cascade, $R_{i,1}+R_{i,2}$, $X_{i,1}+X_{i,2}$ \\
$i_{i,T}^{d},\ i_{i,T}^{q}$ & cascade current in the $dq$ frame \\
$i_{i,\mathrm{IT}}^{dq},\ i_{i,\mathrm{cool}}^{dq},\ i_{i,z}^{dq}$ & IT (UPS rectifier), cooling, and constant-impedance branch currents at the LV bus \\
$v_{i,0}^{d},\ v_{i,0}^{q}$ & transmission-bus (POI) voltage in the $dq$ frame \\
$v_{i,\mathrm{MV}}^{d},\ v_{i,\mathrm{MV}}^{q}$ & medium-voltage bus voltage in the $dq$ frame \\
$v_{i,\mathrm{LV}}^{d},\ v_{i,\mathrm{LV}}^{q}$ & low-voltage bus voltage in the $dq$ frame \\
$V_{i}^{\mathrm{dc}},\ \delta_{i}^{\mathrm{dc}}$ & LV-bus voltage magnitude and angle of data center $i$ \\
$P_{i}^{\mathrm{dc},0},\ Q_{i}^{\mathrm{dc},0}$ & power-flow load of the bus replaced by data center $i$ \\
$P_{i}^{\mathrm{LV}},\ Q_{i}^{\mathrm{LV}}$ & active and reactive power delivered at the LV bus \\
$S_{\mathrm{base}},\ S_{\mathrm{rated}}$ & system and nameplate power bases \\
$Z_{\mathrm{sys}},\ Z_{\mathrm{nameplate}}$ & impedance on the system and nameplate bases \\
\bottomrule
\end{tabularx}

\subsection*{Data-Center IT, Miscellaneous, and Cooling Loads}
\begin{tabularx}{\linewidth}{@{}lX@{}}
\toprule
Symbol & Description \\
\midrule
$P_{i}^{\mathrm{IT}},\ Q_{i}^{\mathrm{IT}}$ & IT-load active and reactive power \\
$P_{i,0}^{\mathrm{IT}}$ & IT active power at the initial operating point ($t=0$) \\
$g_{i}(t)$ & normalized IT/GPU workload profile ($g_i\equiv1$ at nominal utilization) \\
$V_{i}^{\mathrm{fil}}$ & low-pass-filtered LV-bus voltage of the PFC front end \\
$T_{f,\mathrm{PFC}}$ & IT-load PFC voltage-filter time constant \\
$P_{i}^{\mathrm{meas}}$ & measured GPU power-trace segment assigned to data center $i$, shifted to start at $t=0$ \\
$P_{i}^{\mathrm{IT,u}},\ Q_{i}^{\mathrm{IT,u}}$ & IT demand per UPS unit, drawn from its output inverter (pu on $S_n$) \\
$N_{i}^{\mathrm{ups}},\ k_{i}^{\mathrm{ups}}$ & number of identical UPS units and fleet scale $N_{i}^{\mathrm{ups}}S_n/S_{\mathrm{base}}$ \\
$P_{i}^{z},\ Q_{i}^{z}$ & miscellaneous constant-impedance load power \\
$G_{i}^{z},\ B_{i}^{z}$ & equivalent shunt conductance and susceptance \\
$P_{i}^{\mathrm{cool}},\ Q_{i}^{\mathrm{cool}}$ & cooling-load active and reactive power \\
$P_{i}^{\mathrm{dc}},\ Q_{i}^{\mathrm{dc}}$ & composite data-center active and reactive demand \\
$S_{i}^{\mathrm{cool}}$ & cooling-branch (motor) rating; the motor per-unit base \\
$v_{i,ds},\ v_{i,qs}$ & motor stator voltage in the $dq$ frame ($=v_{i,\mathrm{LV}}^{dq}$) \\
$i_{i,ds},\ i_{i,qs},\ i_{i,dr},\ i_{i,qr}$ & motor stator and rotor $dq$ currents \\
$\psi_{i,ds},\ \psi_{i,qs},\ \psi_{i,dr},\ \psi_{i,qr}$ & motor stator and rotor $dq$ flux linkages \\
$R_s,\ R_r$ & motor stator and rotor resistances \\
$X_s,\ X_r,\ X_m$ & stator leakage, rotor leakage, and magnetizing reactances ($\omega_0L_{ls}$, $\omega_0L_{lr}$, $\omega_0L_m$) \\
$\omega_s\ (=\omega_0),\ \omega_{i,r}$ & synchronous and rotor angular speeds \\
$s_i=(\omega_s-\omega_{i,r})/\omega_s$ & motor slip \\
$T_{i,e},\ T_{i,L}$ & electromagnetic and mechanical load torque \\
$H^{m}$ & motor-plus-load inertia constant \\
\bottomrule
\end{tabularx}

\subsection*{Data-Center UPS Rectifier}
\begin{tabularx}{\linewidth}{@{}lX@{}}
\toprule
Symbol & Description \\
\midrule
$v_{i,g}^{d,r},\ v_{i,g}^{q,r}$ & rectifier AC-side voltage in the $dq$ frame \\
$i_{i,g}^{d,r},\ i_{i,g}^{q,r}$ & rectifier AC-side current in the $dq$ frame \\
$v_{i,g}^{q,r,\mathrm{fil}}$ & filtered quadrature-axis voltage \\
$i_{i,g}^{d,r,\mathrm{fil}},\ i_{i,g}^{q,r,\mathrm{fil}}$ & filtered $dq$ currents \\
$i_{i,g,\mathrm{ref}}^{d,r},\ i_{i,g,\mathrm{ref}}^{q,r}$ & $dq$ current references \\
$\theta_{i,\mathrm{PLL}}^{r},\ I_{i,\mathrm{PLL}}^{r},\ \Delta_{i,\mathrm{PLL}}^{r}$ & PLL angle, integrator state, and PI output \\
$\hat{\omega}_{i}^{r}=\omega_0+\dot{\theta}_{i,\mathrm{PLL}}^{r}$ & PLL-estimated angular frequency \\
$v_{i,g}^{d,r},\ v_{i,g}^{q,r}$ & LV-bus voltage projected onto the rectifier PLL frame \\
$P_{i,\mathrm{rect}},\ Q_{i,\mathrm{rect}}$ & rectifier active and reactive power (positive $P$: AC to DC) \\
$P_{i,\mathrm{rect}}^{W}$ & rectifier power delivered to the DC bus, in watts \\
$Q_{i,\mathrm{rect}}^{\mathrm{fil}}$ & filtered rectifier reactive power \\
$Q_{i,\mathrm{ref}}^{r}$ & rectifier reactive-power reference \\
$V_{i,\mathrm{dc}},\ V_{i,\mathrm{dc}}^{\mathrm{fil}}$ & DC-bus voltage and its filtered measurement \\
$V_{i,\mathrm{dc,ref}}$ & DC-bus voltage reference \\
$I_{i,\mathrm{DC}},\ I_{i,\mathrm{load}}$ & rectifier and load DC-side currents \\
$P_{i,\mathrm{load}}$ & DC-side load power \\
$I_{i,\mathrm{Vdc}}^{d,r}$ & DC-bus-voltage-loop integrator state \\
$I_{i,Q}^{q,r}$ & reactive-power-loop integrator state \\
$I_{i,d,\max}^{r}$ & rectifier $d$-axis current limit \\
$I_{i,\mathrm{cc}}^{d,r},\ I_{i,\mathrm{cc}}^{q,r}$ & current-controller integrator states \\
$\epsilon_i^{d,r},\ \epsilon_i^{q,r}$ & current-controller PI outputs \\
$e_{i,\mathrm{rect}}^{d},\ e_{i,\mathrm{rect}}^{q}$ & rectifier voltage commands \\
$R_f^{\mathrm{rect}},\ L_f^{\mathrm{rect}}$ & rectifier filter resistance and inductance \\
$T_{f,\mathrm{PLL}}^{\mathrm{rect}},\ T_{f,\mathrm{PQ}}^{\mathrm{rect}},\ T_{f,\mathrm{ig}}^{\mathrm{rect}}$ & PLL, outer-loop, and current-loop filter time constants \\
$k_{P,\mathrm{PLL}}^{\mathrm{rect}},\ k_{I,\mathrm{PLL}}^{\mathrm{rect}}$ & PLL PI gains \\
$k_{P,\mathrm{Vdc}}^{\mathrm{rect}},\ k_{I,\mathrm{Vdc}}^{\mathrm{rect}}$ & DC-bus-voltage-loop PI gains \\
$k_{P,\mathrm{PQ}}^{\mathrm{rect}},\ k_{I,\mathrm{PQ}}^{\mathrm{rect}}$ & reactive-power-loop PI gains \\
$k_{P,\mathrm{ig}}^{\mathrm{rect}},\ k_{I,\mathrm{ig}}^{\mathrm{rect}}$ & current-controller PI gains \\
$\alpha_{\mathrm{leak}}^{\mathrm{rect}},\ \alpha_{\mathrm{leak,cc}}^{\mathrm{rect}}$ & outer- and inner-loop anti-windup leakage coefficients \\
$V_{\mathrm{base}},I_{\mathrm{base}},Z_{\mathrm{base}},L_{\mathrm{base}}$ & per-unit base quantities \\
$V_{\mathrm{LL,rms}},\ S_n$ & rated line-to-line RMS voltage and apparent power \\
\bottomrule
\end{tabularx}

\subsection*{Data-Center UPS Battery, Converter, and DC Bus}
\begin{tabularx}{\linewidth}{@{}lX@{}}
\toprule
Symbol & Description \\
\midrule
$\mathrm{SOC}_i$ & battery state of charge \\
$I_{i,\mathrm{bat}}$ & battery current (positive on discharge) \\
$E_i$ & battery internal (no-load) voltage \\
$V_{i,\mathrm{bat}}$ & battery terminal voltage \\
$P_{i,\mathrm{bat}}$ & battery power (positive on discharge) \\
$E_{i,0}$ & constant open-circuit voltage \\
$K_i$ & polarization constant \\
$A_i,\ B_i$ & exponential-zone amplitude and inverse capacity \\
$Q_i,\ Q_{i,\mathrm{As}}$ & nominal capacity in ampere-hours and ampere-seconds \\
$R_{i,\mathrm{int}}$ & battery internal ohmic resistance \\
$d_i$ & current-direction-dependent polarization denominator \\
$w_i$ & smooth discharge weight, $\tfrac12(1+\operatorname{sign}_\varepsilon(i_{i,L}))$ \\
$i_{i,L},\ i_{i,H}$ & converter low- and high-voltage-side currents \\
$D_i$ & converter duty cycle; $D_{i,\min},D_{i,\max}$ its limits \\
$D_{i,\mathrm{ff}}$ & feedforward duty cycle \\
$L_{i,\mathrm{dc}},\ R_{i,\mathrm{dc}}$ & converter filter inductance and series resistance \\
$e_{i,\mathrm{Vdc}},\ x_{i,\mathrm{Vdc}}$ & DC-bus-voltage error and integrator state \\
$I_{i,\mathrm{ref}}^{\mathrm{bat}}$ & battery current reference \\
$e_{i,I}$ & inductor-current error \\
$k_{P,\mathrm{Vdc}}^{\mathrm{bi}},\ k_{I,\mathrm{Vdc}}^{\mathrm{bi}}$ & converter DC-bus-voltage-loop PI gains \\
$k_{P,\mathrm{ig}}^{\mathrm{bi}}$ & converter inner current-loop proportional gain \\
$\tau_{i,\mathrm{cl}}$ & closed-loop inductor-current time constant \\
$C_{i,\mathrm{dc}}$ & DC-link capacitance \\
$P_{i,\mathrm{inv}}^{W}$ & output-inverter active power drawn from the DC bus, in watts \\
\bottomrule
\end{tabularx}

\subsection*{Data-Center UPS Output Inverter (Grid-Forming)}
\begin{tabularx}{\linewidth}{@{}lX@{}}
\toprule
Symbol & Description \\
\midrule
$f_{i}^{\mathrm{droop}},\ V_{i}^{\mathrm{droop}}$ & droop frequency and voltage-magnitude references \\
$\theta_{i}^{\mathrm{inv}}$ & inverter angle relative to the synchronous frame \\
$x_{i}^{V}$ & voltage-loop integrator state \\
$v_{i}^{d,\mathrm{inv}},\ v_{i}^{q,\mathrm{inv}}$ & UPS output-bus voltage in the inverter frame \\
$V_{i}^{\mathrm{inv}}$ & UPS output-bus voltage magnitude \\
$i_{i}^{d,\mathrm{inv}},\ i_{i}^{q,\mathrm{inv}}$ & current drawn by the computing load \\
$P_{i,\mathrm{inv}},\ Q_{i,\mathrm{inv}}$ & inverter output active and reactive power (pu on $S_n$) \\
$f_{\mathrm{ref}},\ V_{\mathrm{AC,ord}}$ & frequency reference and AC voltage order \\
$P_{\mathrm{set}},\ Q_{\mathrm{set}}$ & droop power set points \\
$D_{pf},\ D_{qV},\ D_{VV}$ & $P$--$f$, $Q$--$V$, and $V_{\mathrm{AC}}$--$V_{\mathrm{DC}}$ droop gains \\
$k_{P,V}^{\mathrm{ups}},\ k_{I,V}^{\mathrm{ups}}$ & voltage-loop PI gains \\
\bottomrule
\end{tabularx}

\subsection*{Data-Center Bidirectional DC--DC Converter Control}
\begin{tabularx}{\linewidth}{@{}lX@{}}
\toprule
Symbol & Description \\
\midrule
$\tilde{e}_{i}$ & one-sided DC-bus voltage error beyond the dead band \\
$\delta_{\mathrm{db}}$ & DC-bus dead band (per unit of $V_{i,\mathrm{dc,ref}}$) \\
$\alpha^{\mathrm{bi}}$ & outer-loop integrator leakage \\
$I_{i,\mathrm{ch}},\ I_{\mathrm{ch,max}}$ & charger current and its maximum \\
$\phi_i,\ \phi_0$ & charger SOC taper and its value at $\mathrm{SOC}_{i,0}$ \\
$\gamma_i$ & charger gate (off when the DC bus sags) \\
$\Delta_{\mathrm{SOC}}$ & charger taper width \\
\bottomrule
\end{tabularx}

\subsection*{Smooth Operators and Discretization}
\begin{tabularx}{\linewidth}{@{}lX@{}}
\toprule
Symbol & Description \\
\midrule
$\smax(\cdot),\ \smin(\cdot)$ & smooth maximum and minimum \\
$\sclip(x,\underline{x},\bar{x})$ & smooth saturation of $x$ to $[\underline{x},\bar{x}]$ \\
$\ssign(x)$ & smooth sign function \\
$|x|_\varepsilon$ & smooth absolute value \\
$\varepsilon$ & smoothing parameter \\
$h$ & integration step size \\
\bottomrule
\end{tabularx}

\section{Positive Sequence Transmission Network Model (Slow Subsystem)}
\subsection{Synchronous Generator Model}
We model the synchronous generator index $i \in \mathcal{N}_g  = \{1,\dots,N_g\}$ with the classical second-order swing equation in \eqref{eq:syn_gen}.
The model includes a constant internal voltage $E_{i}'$ behind a transient reactance $X_{i}^{d'}$. 
Here $N_g$ is the total number of generators, with the generator $i$ connected to the bus, $j \in \mathcal{N}_b = \{1,\dots,N_b\}$ and $N_b$ the total number of buses.  
In \eqref{eq:syn_gen}, $\delta_i^{g}$ denotes the generator $i$ internal  angle, $\delta_{j}^{b}$ is the voltage angle of bus $j$, $\Delta\omega_i^{g}$ is the per-unit rotor-speed deviation of the generator $i$ on the base $\omega_0 = 2\pi f_0$, and $V_{j}^{b}$ is the bus $j$ voltage magnitude.

\begin{subequations}\label{eq:syn_gen}
{\setlength{\jot}{2pt}
\begin{align}
  \dot{\delta}_i^{g} &= \omega_0\,\Delta \omega_i^{g}, \label{eq:delta_dot}\\
  \dot{\Delta \omega}_i^{g} &= \frac{1}{2H_i^{g}}
  \big( P_{i}^{m} - D_i^{g}\,\Delta \omega_i^{g} - P_{i}^{e} \big), \label{eq:omega_dot}\\
  P_{i}^{e} &= \frac{E_{i}'\,V_{j}^{b}}{X_{i}^{d'}}
  \sin(\delta_{i}^{g} - \delta_{j}^{b}), \label{eq:Pe}\\
  Q_{i}^{e} &= \frac{E_{i}'\,V_{j}^{b}}{X_{i}^{d'}}
  \cos(\delta_{i}^{g} - \delta_{j}^{b})
  - \frac{(V_{j}^{b})^2}{X_{i}^{d'}}. \label{eq:Qe}
\end{align}
}
\end{subequations}

\noindent Also, for generator $i$, \(H_i^{g}\) is the inertia constant, \(D_i^{g}\) is the damping coefficient, \(P_{i}^{m}\) is the mechanical input power (p.u.), and \(P_{i}^{e}, Q_{i}^{e}\) are the electrical active/reactive powers, respectively. This model neglects stator resistance and saliency, treating \(E_{i}'\) as constants.

\subsubsection{Turbine-Governor Model}
We model the governor–turbine aggregate for the generator $i \in \mathcal{N}_g$ as a first-order low-pass filter with time constant, $T_{i}^{\mathrm{gov}}$, driven by a power setpoint,$P_{i}^{\mathrm{ref}}$, as:
\begin{subequations}\label{eq:governor_equations}
\begin{align}
  \dot{P}_{i}^{m} &= \frac{1}{T_{i}^{\mathrm{gov}}}
  \left(-P_{i}^{m} + P_{i}^{\mathrm{ref}} - \frac{\Delta\omega_i^g}{R_i}\right), \label{eq:gov_dyn}\\[3pt]
  \underline{P}_{i}^{m} &\leq P_{i}^{m} \leq \overline{P}_{i}^{m}. \label{eq:gov_limits}
\end{align}
\end{subequations}
In~(\ref{eq:gov_dyn}), $R_i$ represents the droop of the governor and \(\overline{P}_{i}^{m}\) and \(\underline{P}_{i}^{m}\) are the upper and lower bounds on the mechanical power. 
These equations collectively model the dynamic behavior of the generator and its interaction with the turbine governor.
\subsubsection{AC power balance}
To couple the machine dynamics with the network, algebraic equations impose AC power-balance constraints at each bus $i \in \mathcal{N}_b$: 

\begin{subequations}\label{eq:ac_pf}
\begin{align}
  P_{i}^{g}-P_{i}^{d}
  &= V_i^{b} \sum_{k} V_k^{b}\!\left[\,G_{ik}\cos(\delta_{ik}^{b})+B_{ik}\sin(\delta_{ik}^{b})\right], \label{eq:pf_P_ik}\\[3pt]
  Q_{i}^{g}-Q_{i}^{d}
  &= V_i^{b} \sum_{k} V_k^{b}\!\left[\,G_{ik}\sin(\delta_{ik}^{b})-B_{ik}\cos(\delta_{ik}^{b})\right]. \label{eq:pf_Q_ik}
\end{align}
\end{subequations}

\noindent where, \(G_{ik}, B_{ik}\) is the conductance and susceptance for the line between nodes are $i$ and $k$ while \(\delta_{ik}^{b} = \delta_{i}^{b} -\delta_{k}^{b}\) The generator aggregates at bus \(i\) are \(P_{i}^{g}=\sum_{j\in\mathcal{N}_g} P_{j}^{e}\) and \(Q_{i}^{g}=\sum_{j\in\mathcal{N}_g} Q_{j}^{e}\). We represent each load $i \in \mathcal{N}_{l} = \{1,\dots,N_l\}$, where $N_l$ is the total number of loads, with a ZIP model\footnote{In this work, active power load is converted to the constant load, and reactive power load is converted to the constant impedance load}.
\section{Fast Sub-Systems}
\subsection{Grid-Following Inverter Model}\label{sec:gfl}
Inverter-based resources (IBRs) typically interface with the power system through voltage-source converter (VSC) technology. While the underlying converter hardware remains similar across applications, the dynamic behavior of an IBR depends critically on the selected control architecture. In the subsequent section, we focus on grid-following control strategies and describe their internal control structure and dynamic characteristics.

\begin{figure}[H]
    \centering
\includegraphics[width=0.7\columnwidth,keepaspectratio]{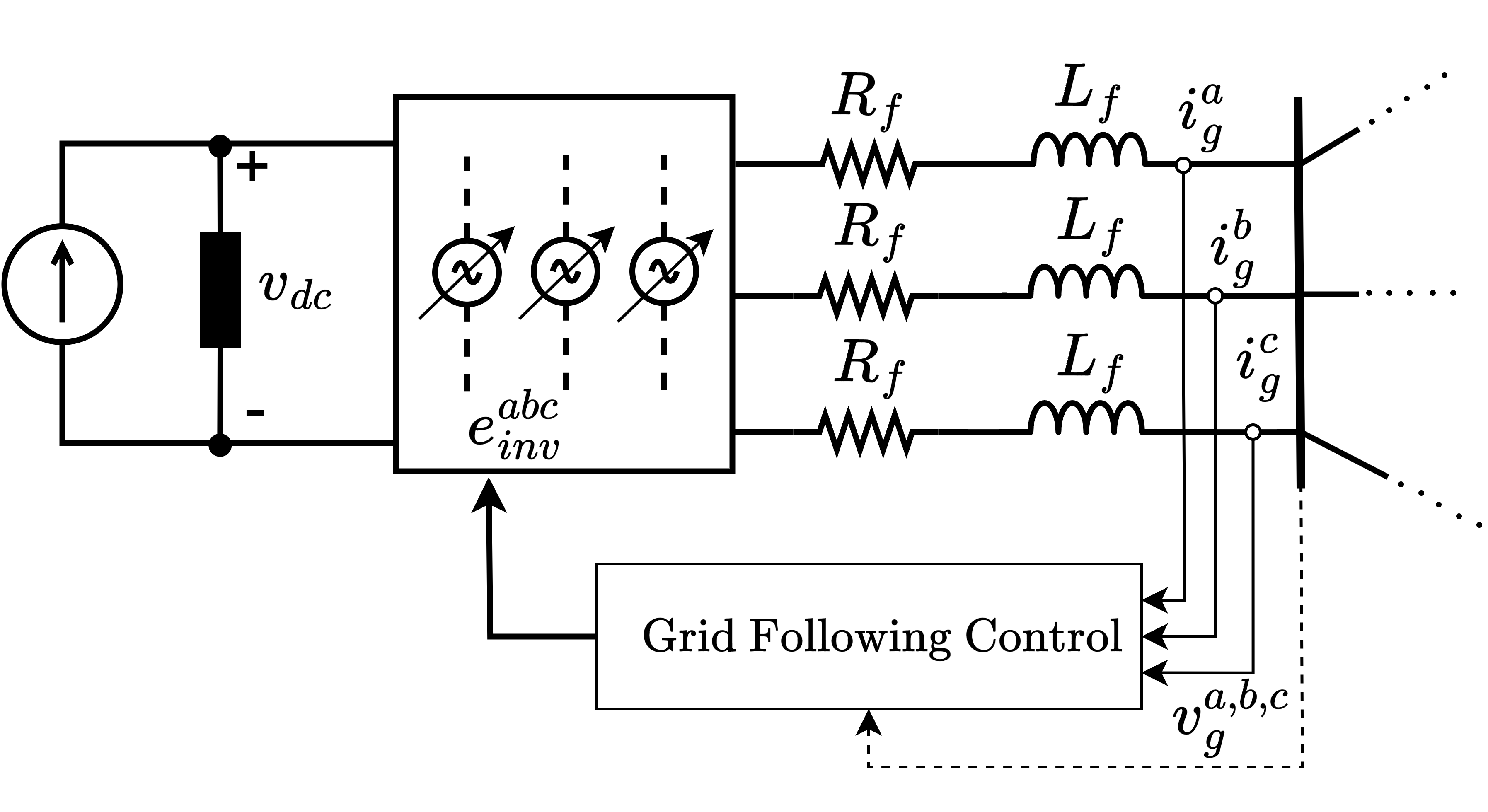}
    \caption{Grid-connected inverter equivalent circuit and controller.}
    \label{fig:inv_model}
\end{figure}

For the transmission network, we adopt a positive-sequence representation and solve the network dynamics in the electromechanical/phasor domain, as described in the previous section. In contrast, for inverter-based resources, we explicitly model the instantaneous terminal voltage waveforms to capture fast dynamics in the control loops. Figure~\ref{fig:inv_model} illustrates the inverter configuration, where a voltage-source converter connects to the network bus through an $R_f^{\mathrm{inv}}$–$L_f^{\mathrm{inv}}$ output filter.

We introduce two standard modeling assumptions. First, we assume the DC-link voltage remains constant over the simulation horizon. Second, we neglect switching dynamics and represent the converter as an ideal controllable voltage source, which is a well-established assumption in power electronics modeling~\cite{erickson2007fundamentals}, \cite{yazdani2010voltage}. Under these assumptions, the control outputs $e_{i,\mathrm{inv}}^{dq}$ directly command the inverter terminal voltage in the synchronous reference frame. 

\begin{figure}[H]
    \centering
\includegraphics[width=0.7\columnwidth,keepaspectratio]{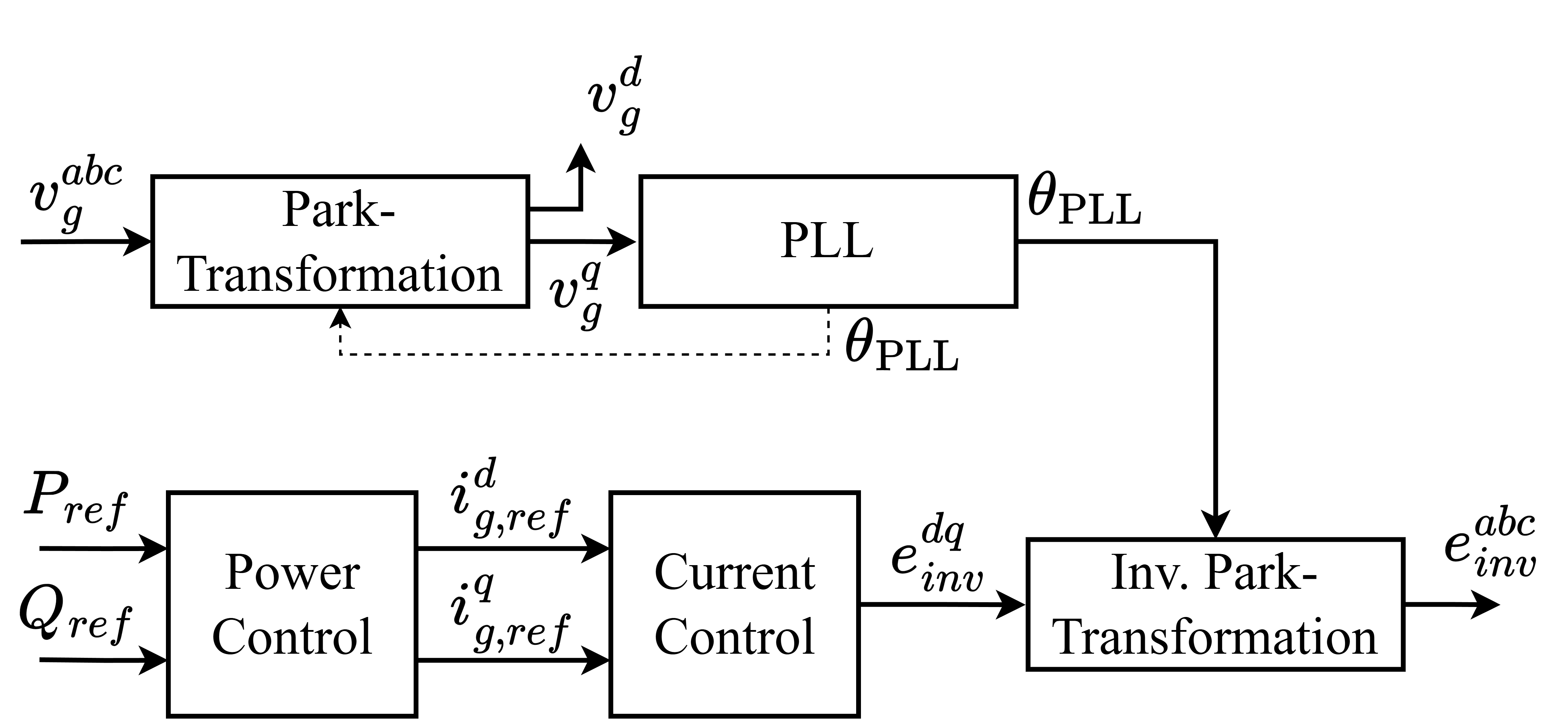}
    \caption{Grid-following inverter control block diagram.}
    \label{fig:inv_control}
\end{figure}
The grid-following inverter control architecture consists of three main components\footnote{We carry out all inverter control and electrical calculations in the per-unit system}: a phase-locked loop (PLL), a power controller, and a current controller. The PLL synchronizes the inverter reference frame with the grid voltage, the power controller regulates active and reactive power outputs, and the current controller generates voltage commands to track the desired current references.
\subsubsection{Phase-Locked Loop (PLL)}

We model the phase-locked loop (PLL) of the inverter index 
$i \in \mathcal{N}_{\mathrm{inv}} = \{1,\dots,N_{\mathrm{inv}}\}$ 
to estimate the grid voltage phase angle required for synchronous reference-frame control. 
The PLL transforms the measured three-phase grid voltages into the rotating $dq$ frame 
using the estimated PLL angle $\theta_{i,\mathrm{PLL}}(t)$ via the Park transformation 
and enforces phase alignment by regulating the quadrature-axis voltage to zero. 
To attenuate measurement noise, we filter the quadrature-axis voltage using a first-order low-pass filter.
\begin{subequations}\label{eq:pll_dynamics}
\begin{align}
\frac{d v_{i,g}^{q,\mathrm{fil}}(t)}{dt}
&=
\frac{1}{T_{f,\mathrm{PLL}}^{\mathrm{inv}}}
\big(
v_{i,g}^{q}(t)
-
v_{i,g}^{q,\mathrm{fil}}(t)
\big), \label{eq:pll_lpf} \\[4pt]
\frac{d I_{i,\mathrm{PLL}}(t)}{dt}
&=
k_{I,\mathrm{PLL}}^{\mathrm{inv}}\,
v_{i,g}^{q,\mathrm{fil}}(t), \label{eq:pll_integrator} \\[4pt]
\Delta_{i,\mathrm{PLL}}(t)
&=
k_{P,\mathrm{PLL}}^{\mathrm{inv}}\,
v_{i,g}^{q,\mathrm{fil}}(t)
+
I_{i,\mathrm{PLL}}(t), \label{eq:pll_pi} \\[4pt]
%
%
\frac{d \theta_{i,\mathrm{PLL}}(t)}{dt}
&=
\omega_0\, \Delta_{i,\mathrm{PLL}}(t). \label{eq:pll_angle}
\end{align}
\end{subequations}

In PLL, the PI controller \eqref{eq:pll_pi} generates a frequency deviation proportional to the voltage phase error. Integrating this deviation yields the estimated grid angle, which synchronizes the inverter reference frame with the grid.

\subsubsection{Power Controller}

We model the active and reactive power controller of the inverter index 
$i \in \mathcal{N}_{\mathrm{inv}} = \{1,\dots,N_{\mathrm{inv}}\}$ 
as the upper-level control that generates current references to achieve the desired power injection at the point of common coupling (PCC). 
Using the synchronous reference frame, we compute the instantaneous active and reactive power \footnote{a positive quadrature-axis current $i_{i,g}^{q}$ produces positive reactive power $Q_{i,g}$} as
\begin{subequations}\label{eq:PQ_dq}
\begin{align}
    P_{i,g}(t) &= \frac{3}{2} \Big( v_{i,g}^{d}(t)\, i_{i,g}^{d}(t)
                              + v_{i,g}^{q}(t)\, i_{i,g}^{q}(t) \Big), \label{eq:P_dq} \\[2mm]
    Q_{i,g}(t) &= \frac{3}{2} \Big( v_{i,g}^{d}(t)\, i_{i,g}^{q}(t)
                              - v_{i,g}^{q}(t)\, i_{i,g}^{d}(t) \Big). \label{eq:Q_dq}
\end{align}
\end{subequations}

We filter the instantaneous power using a first-order low-pass filter with a time constant $T_{f,\mathrm{PQ}}^{\mathrm{inv}}$, The time constant of the filter provides a degree of freedom for tuning the controller gains\footnote{Symmetrical Optimum criterion: The method "symmetrically" places the crossover frequency of the open-loop system to maximize the phase margin, which in turn ensures the controller is fast and also stable}.
\begin{subequations}\label{eq:PQ_fil}
\begin{align}
\frac{d P_{i,g}^{\mathrm{fil}}(t)}{dt} &= \frac{1}{T_{f,\mathrm{PQ}}^{\mathrm{inv}}}
\big( P_{i,g}(t) - P_{i,g}^{\mathrm{fil}}(t) \big), \label{eq:gfl_Pfil}\\
\frac{d Q_{i,g}^{\mathrm{fil}}(t)}{dt} &= \frac{1}{T_{f,\mathrm{PQ}}^{\mathrm{inv}}}
\big( Q_{i,g}(t) - Q_{i,g}^{\mathrm{fil}}(t) \big). \label{eq:gfl_Qfil}
\end{align}
\end{subequations}

We regulate the filtered active and reactive power using proportional–integral (PI) controllers to generate the current references for the inner current control loop. 
The controller dynamics are given by
\begin{subequations}\label{eq:ig_dref}
\begin{align}
\frac{d I_{i,P}^{d}(t)}{dt} &= k_{I,\mathrm{PQ}}^{\mathrm{inv}}
\big( P_{i,\mathrm{ref}}(t) - P_{i,g}^{\mathrm{fil}}(t) \big), \label{eq:gfl_IP}\\
i_{i,g,\mathrm{ref}}^{d}(t) &=
k_{P,\mathrm{PQ}}^{\mathrm{inv}}
\big( P_{i,\mathrm{ref}}(t) - P_{i,g}^{\mathrm{fil}}(t) \big)
+ I_{i,P}^{d}(t), \label{eq:gfl_idref}\\
\frac{d I_{i,Q}^{q}(t)}{dt} &= k_{I,\mathrm{PQ}}^{\mathrm{inv}}
\big( Q_{i,\mathrm{ref}}(t) - Q_{i,g}^{\mathrm{fil}}(t) \big), \label{eq:gfl_IQ}\\
i_{i,g,\mathrm{ref}}^{q}(t) &=
k_{P,\mathrm{PQ}}^{\mathrm{inv}}
\big( Q_{i,\mathrm{ref}}(t) - Q_{i,g}^{\mathrm{fil}}(t) \big)
+ I_{i,Q}^{q}(t). \label{eq:gfl_iqref}
\end{align}
\end{subequations}

\subsubsection{Current Controller}

We model the current controller of the inverter index 
$i \in \mathcal{N}_{\mathrm{inv}} = \{1,\dots,N_{\mathrm{inv}}\}$ 
as the inner-loop controller that regulates the grid currents in the synchronous reference frame. 
To attenuate noise and tune the controller gains, we filter the measured currents using a first-order low-pass filter with time constant $T_{f,\mathrm{ig}}^{\mathrm{inv}}$,
\begin{subequations}\label{eq:ig_fil}
\begin{align}
    \frac{d i_{i,g}^{d,\mathrm{fil}}(t)}{dt} &=
    \frac{1}{T_{f,\mathrm{ig}}^{\mathrm{inv}}}
    \big(i_{i,g}^{d}(t) - i_{i,g}^{d,\mathrm{fil}}(t) \big), \label{eq:gfl_ifil_d}\\
    \frac{d i_{i,g}^{q,\mathrm{fil}}(t)}{dt} &=
    \frac{1}{T_{f,\mathrm{ig}}^{\mathrm{inv}}}
    \big(i_{i,g}^{q}(t) - i_{i,g}^{q,\mathrm{fil}}(t) \big). \label{eq:gfl_ifil_q}
\end{align}    
\end{subequations}

We regulate the filtered currents to track the references $(i_{i,g,\mathrm{ref}}^{d}(t), i_{i,g,\mathrm{ref}}^{q}(t))$ using PI controllers,
\begin{subequations}\label{eq:PI_cc}
\begin{align}
    \frac{d I_{i,\mathrm{cc}}^{d}(t)}{dt} &=
    k_{I,\mathrm{ig}}^{\mathrm{inv}}
    \big(i_{i,g,\mathrm{ref}}^{d}(t) - i_{i,g}^{d,\mathrm{fil}}(t) \big), \label{eq:gfl_cc_Id}\\
    \epsilon_{i}^{d}(t) &=
    k_{P,\mathrm{ig}}^{\mathrm{inv}}
    \big(i_{i,g,\mathrm{ref}}^{d}(t) - i_{i,g}^{d,\mathrm{fil}}(t) \big)
    + I_{i,\mathrm{cc}}^{d}(t), \label{eq:gfl_cc_ed}\\
    \frac{d I_{i,\mathrm{cc}}^{q}(t)}{dt} &=
    k_{I,\mathrm{ig}}^{\mathrm{inv}}
    \big(i_{i,g,\mathrm{ref}}^{q}(t) - i_{i,g}^{q,\mathrm{fil}}(t) \big), \label{eq:gfl_cc_Iq}\\
    \epsilon_{i}^{q}(t) &=
    k_{P,\mathrm{ig}}^{\mathrm{inv}}
    \big(i_{i,g,\mathrm{ref}}^{q}(t) - i_{i,g}^{q,\mathrm{fil}}(t) \big)
    + I_{i,\mathrm{cc}}^{q}(t). \label{eq:gfl_cc_eq}
\end{align}
\end{subequations}

We generate the inverter voltage commands by augmenting the PI outputs with decoupling terms and grid-voltage feedforward.
\begin{subequations}\label{eq:e_AC_dq}
\begin{align}
    e_{i,\mathrm{inv}}^{d}(t) &=
    \epsilon_{i}^{d}(t)
    - \hat{\omega}_{i}(t)\, L_f^{\mathrm{inv}}\, i_{i,g}^{q}(t)
    + v_{i,g}^{d}(t), \label{eq:gfl_ed}\\
    e_{i,\mathrm{inv}}^{q}(t) &=
    \epsilon_{i}^{q}(t)
    + \hat{\omega}_{i}(t)\, L_f^{\mathrm{inv}}\, i_{i,g}^{d}(t)
    + v_{i,g}^{q}(t). \label{eq:gfl_eq}
\end{align}
\end{subequations}

Equivalently, the $dq$-frame filter dynamics satisfy the continuous-time relationships
\begin{subequations}\label{eq:KVL_equations}
{\footnotesize
\begin{align}
L_f^{\mathrm{inv}} \frac{d i_{i,g}^{d}(t)}{dt} &=
-R_f^{\mathrm{inv}} i_{i,g}^{d}(t)
+ \hat{\omega}_{i}(t) L_f^{\mathrm{inv}} i_{i,g}^{q}(t)
+ e_{i,\mathrm{inv}}^{d}(t)
- v_{i,g}^{d}(t) \label{eq:gfl_kvl_d}\\
L_f^{\mathrm{inv}} \frac{d i_{i,g}^{q}(t)}{dt} &=
-R_f^{\mathrm{inv}} i_{i,g}^{q}(t)
- \hat{\omega}_{i}(t) L_f^{\mathrm{inv}} i_{i,g}^{d}(t)
+ e_{i,\mathrm{inv}}^{q}(t)
- v_{i,g}^{q}(t) \label{eq:gfl_kvl_q}
\end{align}
}
\end{subequations}

\subsection{Data Center Model}
\begin{figure}[H]
    \centering
\includegraphics[width=0.8\columnwidth,keepaspectratio]{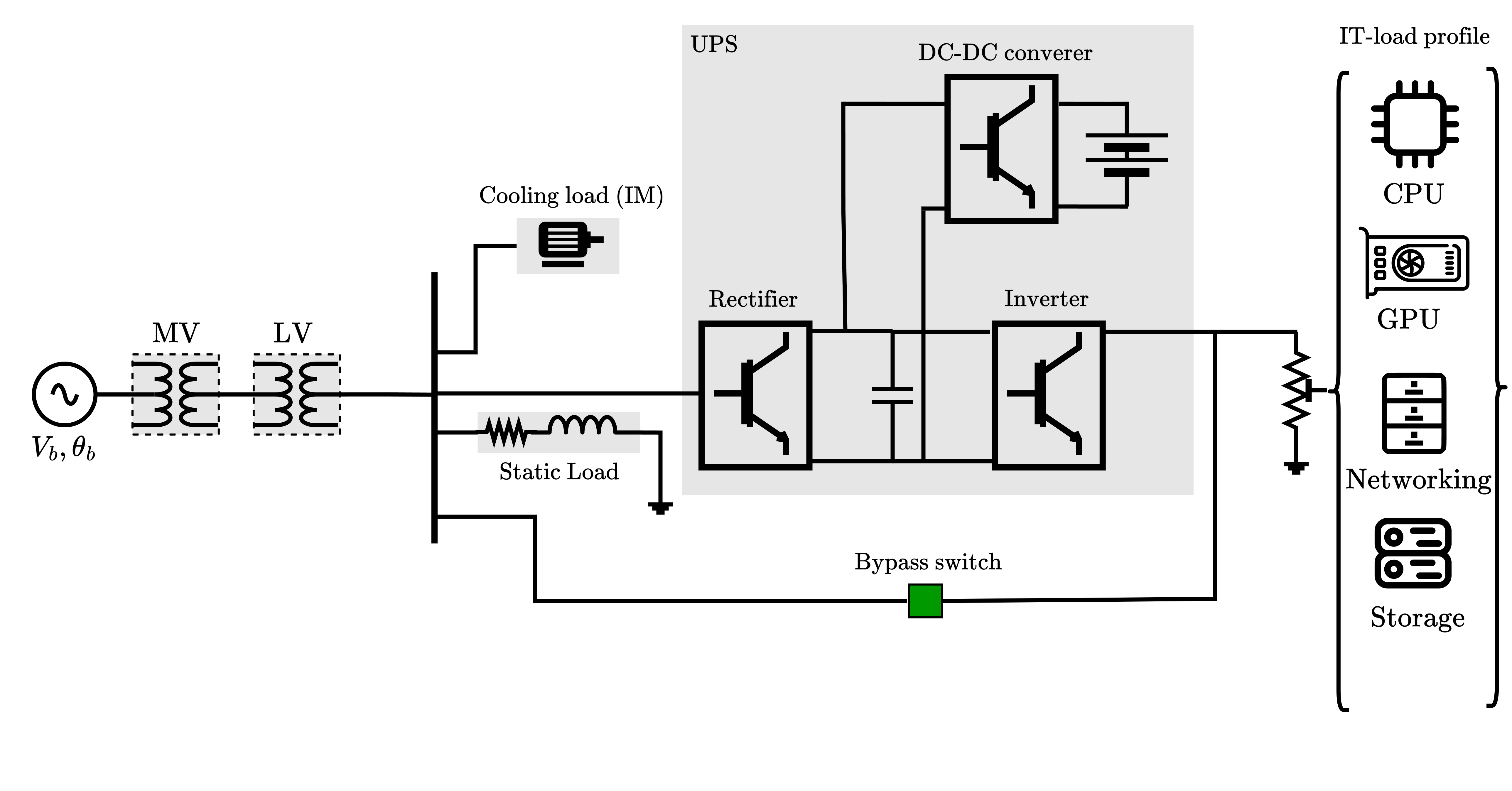}
    \caption{Double-conversion UPS-based AI data center design.}
    \label{fig:dc}
\end{figure}

We model the double-conversion UPS-based AI data center as connected to
the transmission bus, characterized by the voltage magnitude $V_{j}^{b}$ and phase angle $\delta_{j}^{b}$, through two cascaded step-down transformers.
Its low-voltage (LV) bus supplies three parallel branches: a
UPS-protected IT load, an induction-motor cooling load, and a residual constant-impedance load as shown in the Fig.\ref{fig:dc}.

\subsubsection{Transformer Model}\label{sec:xfmr}
We model the step-down transformers of data center
$i\in\mathcal{N}_{\mathrm{dc}}=\{1,\ldots,N_{\mathrm{dc}}\}$, which connect transmission bus $j$, with voltage $V_{j}^{b}\angle\delta_{j}^{b}$, to the $480$~V data-center bus through T1 (HV/MV) and T2 (MV/LV) in cascade. At nominal tap, with per-zone voltage bases equal to the nominal winding voltages, the ideal turns ratio equals $1$~pu, so we represent each transformer $k\in\{1,2\}$ by its series leakage impedance $R_{i,k}+jX_{i,k}$. We neglect the magnetizing branch (about $1\%$ of rated current) and the Y/$\Delta$ phase shift, and we refer the nameplate leakage to the system base via $Z_{\mathrm{sys}}=Z_{\mathrm{nameplate}}\,(S_{\mathrm{base}}/S_{\mathrm{rated}})$. Because T1 and T2 carry the same current $(i_{i,T}^{d},i_{i,T}^{q})$, we lump them into $R_{i,T}=R_{i,1}+R_{i,2}$ and $X_{i,T}=X_{i,1}+X_{i,2}$. In the synchronous $dq$ frame rotating at $\omega_0$, the cascade current obeys
\begin{subequations}\label{eq:xfmr_rl}
\begin{align}
  \frac{X_{i,T}}{\omega_0}\frac{d i_{i,T}^{d}(t)}{dt} &= v_{i,0}^{d}(t)-v_{i,\mathrm{LV}}^{d}(t)-R_{i,T}\,i_{i,T}^{d}(t)+X_{i,T}\,i_{i,T}^{q}(t), \label{eq:xfmr_rl_d}\\
  \frac{X_{i,T}}{\omega_0}\frac{d i_{i,T}^{q}(t)}{dt} &= v_{i,0}^{q}(t)-v_{i,\mathrm{LV}}^{q}(t)-R_{i,T}\,i_{i,T}^{q}(t)-X_{i,T}\,i_{i,T}^{d}(t), \label{eq:xfmr_rl_q}
\end{align}
\end{subequations}
where $v_{i,0}^{d}=V_{j}^{b}\cos\delta_{j}^{b}$ and $v_{i,0}^{q}=V_{j}^{b}\sin\delta_{j}^{b}$ denote the transmission-bus voltage, and the $\pm X_{i,T}\,i_{i,T}^{q/d}$ terms denote the rotating-frame cross-coupling, as in the converter filters of Section~\ref{sec:gfl}. We determine the LV-bus voltage $(v_{i,\mathrm{LV}}^{d},v_{i,\mathrm{LV}}^{q})$ algebraically from the current balance at the LV bus, where the cascade current splits into the IT (UPS rectifier), cooling, and constant-impedance branches:
\begin{subequations}\label{eq:xfmr_kcl}
\begin{align}
  i_{i,T}^{d}(t) &= i_{i,\mathrm{IT}}^{d}(t)+i_{i,\mathrm{cool}}^{d}(t)+i_{i,z}^{d}(t), \label{eq:xfmr_kcl_d}\\
  i_{i,T}^{q}(t) &= i_{i,\mathrm{IT}}^{q}(t)+i_{i,\mathrm{cool}}^{q}(t)+i_{i,z}^{q}(t). \label{eq:xfmr_kcl_q}
\end{align}
\end{subequations}
We recover the MV-bus voltage from the leakage drop across T1,
\begin{subequations}\label{eq:xfmr_mv}
\begin{align}
  v_{i,\mathrm{MV}}^{d}(t) &= v_{i,0}^{d}(t)-R_{i,1}\,i_{i,T}^{d}(t)-\frac{X_{i,1}}{\omega_0}\frac{d i_{i,T}^{d}(t)}{dt}+X_{i,1}\,i_{i,T}^{q}(t), \\
  v_{i,\mathrm{MV}}^{q}(t) &= v_{i,0}^{q}(t)-R_{i,1}\,i_{i,T}^{q}(t)-\frac{X_{i,1}}{\omega_0}\frac{d i_{i,T}^{q}(t)}{dt}-X_{i,1}\,i_{i,T}^{d}(t).
\end{align}
\end{subequations}
We compute the data-center bus voltage magnitude and the power that data center $i$ draws at its point of interconnection (POI), which includes the transformer losses, as
\begin{subequations}\label{eq:xfmr_out}
\begin{align}
  V_{i}^{\mathrm{dc}}(t) &= \sqrt{\big(v_{i,\mathrm{LV}}^{d}(t)\big)^2+\big(v_{i,\mathrm{LV}}^{q}(t)\big)^2}, \\
  P_{i}^{\mathrm{dc}}(t) &= v_{i,0}^{d}(t)\,i_{i,T}^{d}(t)+v_{i,0}^{q}(t)\,i_{i,T}^{q}(t), \qquad
  Q_{i}^{\mathrm{dc}}(t) = v_{i,0}^{q}(t)\,i_{i,T}^{d}(t)-v_{i,0}^{d}(t)\,i_{i,T}^{q}(t).
\end{align}
\end{subequations}
We initialize the cascade at the power-flow current of the replaced load $P_{i}^{\mathrm{dc},0}+jQ_{i}^{\mathrm{dc},0}$,
\begin{equation}\label{eq:xfmr_ic}
  i_{i,T}^{d}(0) = \frac{P_{i}^{\mathrm{dc},0}v_{i,0}^{d}+Q_{i}^{\mathrm{dc},0}v_{i,0}^{q}}{(V_{j}^{b})^{2}}, \qquad
  i_{i,T}^{q}(0) = \frac{P_{i}^{\mathrm{dc},0}v_{i,0}^{q}-Q_{i}^{\mathrm{dc},0}v_{i,0}^{d}}{(V_{j}^{b})^{2}},
\end{equation}
so that the data center starts at its power-flow operating point with $di_{i,T}^{dq}/dt=0$.
The transformer model therefore has two differential states ($i_{i,T}^{d},i_{i,T}^{q}$), which \eqref{eq:xfmr_rl_d}--\eqref{eq:xfmr_rl_q} determine, and two algebraic variables ($v_{i,\mathrm{LV}}^{d},v_{i,\mathrm{LV}}^{q}$), which the current balance \eqref{eq:xfmr_kcl_d}--\eqref{eq:xfmr_kcl_q} determines. These four equations close the model for a given transmission-bus voltage $(v_{i,0}^{d},v_{i,0}^{q})$ and the branch currents of Sections~\ref{sec:it}--\ref{sec:cooling}; \eqref{eq:xfmr_mv} and \eqref{eq:xfmr_out} only evaluate outputs.

\subsubsection{IT Load Model}\label{sec:it}
We model the aggregate IT load of data center
$i\in\mathcal{N}_{\mathrm{dc}}=\{1,\ldots,N_{\mathrm{dc}}\}$ with either a variable-resistor model or a constant-power model. Server power-supply units use power-factor-corrected (PFC) rectifier front ends that draw real power at unity power factor; we do not model their internal switching dynamics.

\emph{Variable-resistor model:} We filter the LV-bus voltage with time constant $T_{f,\mathrm{PFC}}$, which stands in for the PFC's DC-link regulation, and re-adjust the load conductance so that the drawn power tracks the workload-scaled reference $P_{i,0}^{\mathrm{IT}}g_{i}(t)$:
\begin{subequations}\label{eq:it_load}
\begin{align}
  T_{f,\mathrm{PFC}}\,\frac{dV_{i}^{\mathrm{fil}}(t)}{dt} &= V_{i}^{\mathrm{dc}}(t)-V_{i}^{\mathrm{fil}}(t), \label{eq:it_vfil}\\
  P_{i}^{\mathrm{IT}}(t) &= P_{i,0}^{\mathrm{IT}}\,g_{i}(t)\left(\frac{V_{i}^{\mathrm{dc}}(t)}{V_{i}^{\mathrm{fil}}(t)}\right)^{2}, \qquad Q_{i}^{\mathrm{IT}}(t)=0. \label{eq:it_PQ}
\end{align}
\end{subequations}
The load therefore acts as a constant impedance during fast voltage changes and recovers constant-power behavior as $V_{i}^{\mathrm{fil}}$ settles to $V_{i}^{\mathrm{dc}}$.

\emph{Constant-power model:} We place the aggregate servers behind the UPS output inverter (Section~\ref{sec:ups}), which isolates them from the LV-bus voltage, and represent them as a constant-power load that follows a measured workload profile:
\begin{subequations}\label{eq:it_cp}
\begin{align}
  P_{i}^{\mathrm{IT}}(t) &= P_{i,0}^{\mathrm{IT}}\,g_{i}(t), \qquad Q_{i}^{\mathrm{IT}}(t)=0, \label{eq:it_cp_PQ}\\
  g_{i}(t) &= \frac{P_{i}^{\mathrm{meas}}(t)}{P_{i}^{\mathrm{meas}}(0)}, \label{eq:it_cp_g}
\end{align}
\end{subequations}
where $P_{i,0}^{\mathrm{IT}}$ denotes the IT active power of data center~$i$ at the initial operating point, which we set to the IT share of its initial LV-bus demand, $P_{i,0}^{\mathrm{IT}}=0.6\,P_{i}^{\mathrm{LV}}(0)$; $P_{i,0}^{\mathrm{IT}}$ sets the size of the IT load, while the dimensionless profile $g_{i}(t)$ sets its shape. $P_{i}^{\mathrm{meas}}$ denotes the segment of the measured GPU power trace of the NLR AI-inference dataset~\cite{vercellino2026genai}
assigned to data center~$i$, shifted to start at $t=0$. We normalize the segment to its first sample so that $g_{i}(0)=1$ the data center draws exactly its power-flow load at $t=0$; we interpolate between the samples with linear interpolation. The servers draw this demand from the UPS output inverters (Section~\ref{sec:ups_inv}). Each UPS unit, rated $S_n$, serves the demand $P_{i,\mathrm{load}}$ at $g_i=1$, so that, per unit on its own rating, the computing load behind each inverter draws
\begin{equation}\label{eq:it_ref}
  P_{i}^{\mathrm{IT,u}}(t) = \frac{P_{i,\mathrm{load}}}{S_n}\,g_{i}(t), \qquad Q_{i}^{\mathrm{IT,u}}(t)=0,
\end{equation}
so that the UPS units of data center~$i$ together deliver $P_{i}^{\mathrm{IT}}(t)$.
In this work, we use the constant-power model \eqref{eq:it_cp}--\eqref{eq:it_ref} in all simulations.

\subsubsection{Miscellaneous Data-Center Load}
\label{sec:misc_load}

We model the miscellaneous load of data center
$i\in\mathcal{N}_{\mathrm{dc}}=\{1,\ldots,N_{\mathrm{dc}}\}$, which accounts for lighting and other auxiliary electrical loads, as a constant impedance with conductance $G_{i}^{z}$ and susceptance $B_{i}^{z}$ connected to the LV bus. In the synchronous $dq$ frame, the branch draws the current
\begin{subequations}\label{eq:misc_load}
\begin{align}
  i_{i,z}^{d}(t) &= G_{i}^{z}\,v_{i,\mathrm{LV}}^{d}(t)+B_{i}^{z}\,v_{i,\mathrm{LV}}^{q}(t), \label{eq:misc_id}\\
  i_{i,z}^{q}(t) &= G_{i}^{z}\,v_{i,\mathrm{LV}}^{q}(t)-B_{i}^{z}\,v_{i,\mathrm{LV}}^{d}(t), \label{eq:misc_iq}
\end{align}
\end{subequations}
which enters the LV-bus current balance~\eqref{eq:xfmr_kcl}, and consumes the power
\begin{subequations}\label{eq:misc_PQ}
\begin{align}
  P_{i}^{z}(t) &= G_{i}^{z}\big(V_{i}^{\mathrm{dc}}(t)\big)^{2}, \label{eq:misc_P}\\
  Q_{i}^{z}(t) &= B_{i}^{z}\big(V_{i}^{\mathrm{dc}}(t)\big)^{2}, \label{eq:misc_Q}
\end{align}
\end{subequations}
where $V_{i}^{\mathrm{dc}}$ denotes the LV-bus voltage magnitude of~\eqref{eq:xfmr_out}. We set $G_{i}^{z}$ so that the branch draws $10\%$ of the initial LV-bus active power, and we solve $B_{i}^{z}$ at $t=0$ so that the LV bus absorbs exactly the power-flow reactive load. The miscellaneous load therefore has no differential state and four algebraic variables ($i_{i,z}^{d},i_{i,z}^{q},P_{i}^{z},Q_{i}^{z}$), which \eqref{eq:misc_id}--\eqref{eq:misc_iq} and \eqref{eq:misc_P}--\eqref{eq:misc_Q} determine for a given LV-bus voltage $(v_{i,\mathrm{LV}}^{d},v_{i,\mathrm{LV}}^{q})$.

\subsubsection{Cooling Load}\label{sec:cooling}
We model the aggregate cooling plant of data center
$i\in\mathcal{N}_{\mathrm{dc}}=\{1,\ldots,N_{\mathrm{dc}}\}$ as a three-phase squirrel-cage induction motor~\cite{7781182} connected to the LV bus. Cooling typically accounts for about $30\%$ of the total power consumption of a standalone facility~\cite{jimenez2025data}. We formulate the motor in per unit on the cooling-branch rating $S_{i}^{\mathrm{cool}}$, in the synchronous $dq$ frame rotating at $\omega_s=\omega_0$ ($1$~pu), and we describe the rotor speed through the slip $s_i=(\omega_s-\omega_{i,r})/\omega_s$. The stator sees the LV-bus voltage,
\begin{equation}\label{eq:im_vs}
  v_{i,ds}(t)=v_{i,\mathrm{LV}}^{d}(t), \qquad v_{i,qs}(t)=v_{i,\mathrm{LV}}^{q}(t).
\end{equation}
We retain the stator and rotor flux dynamics to capture the electrical transient response of the motor. Because the squirrel-cage rotor windings are short-circuited ($v_{i,dr}=v_{i,qr}=0$), the stator and rotor voltage equations become
\begin{subequations}\label{eq:im_voltage}
\begin{align}
  \frac{1}{\omega_0}\frac{d\psi_{i,ds}(t)}{dt} &= v_{i,ds}(t)-R_s\,i_{i,ds}(t)+\psi_{i,qs}(t), \label{eq:im_v_ds}\\
  \frac{1}{\omega_0}\frac{d\psi_{i,qs}(t)}{dt} &= v_{i,qs}(t)-R_s\,i_{i,qs}(t)-\psi_{i,ds}(t), \\
  \frac{1}{\omega_0}\frac{d\psi_{i,dr}(t)}{dt} &= -R_r\,i_{i,dr}(t)+s_i(t)\,\psi_{i,qr}(t), \\
  \frac{1}{\omega_0}\frac{d\psi_{i,qr}(t)}{dt} &= -R_r\,i_{i,qr}(t)-s_i(t)\,\psi_{i,dr}(t). \label{eq:im_v_qr}
\end{align}
\end{subequations}
The flux linkages determine the four $dq$ currents algebraically,
\begin{subequations}\label{eq:im_flux}
\begin{align}
  \psi_{i,ds}(t) &= (X_s+X_m)\,i_{i,ds}(t)+X_m\,i_{i,dr}(t), \label{eq:im_f_ds}\\
  \psi_{i,qs}(t) &= (X_s+X_m)\,i_{i,qs}(t)+X_m\,i_{i,qr}(t), \\
  \psi_{i,dr}(t) &= (X_r+X_m)\,i_{i,dr}(t)+X_m\,i_{i,ds}(t), \\
  \psi_{i,qr}(t) &= (X_r+X_m)\,i_{i,qr}(t)+X_m\,i_{i,qs}(t), \label{eq:im_f_qr}
\end{align}
\end{subequations}
where $R_s$ and $R_r$ denote the stator and rotor resistances, $X_s=\omega_0L_{ls}$ and $X_r=\omega_0L_{lr}$ the leakage reactances, and $X_m=\omega_0L_m$ the magnetizing reactance. The electromagnetic torque drives the slip dynamics,
\begin{subequations}\label{eq:im_mech}
\begin{align}
  T_{i,e}(t) &= X_m\big(i_{i,qs}(t)\,i_{i,dr}(t)-i_{i,ds}(t)\,i_{i,qr}(t)\big), \label{eq:im_te}\\
  2H^{m}\,\frac{ds_i(t)}{dt} &= T_{i,L}-T_{i,e}(t), \label{eq:im_slip}
\end{align}
\end{subequations}
where $H^{m}$ denotes the motor-plus-load inertia constant and $T_{i,L}$ the constant mechanical load torque, which we set to $T_{i,e}$ at the initial slip $s_0$; we neglect mechanical damping. We compute the cooling-load powers and the current the motor draws from the LV bus, both on the system base, as
\begin{subequations}\label{eq:im_out}
\begin{align}
  P_{i}^{\mathrm{cool}}(t) &= \frac{S_{i}^{\mathrm{cool}}}{S_{\mathrm{base}}}\big(v_{i,ds}\,i_{i,ds}+v_{i,qs}\,i_{i,qs}\big), \qquad
  Q_{i}^{\mathrm{cool}}(t) = \frac{S_{i}^{\mathrm{cool}}}{S_{\mathrm{base}}}\big(v_{i,qs}\,i_{i,ds}-v_{i,ds}\,i_{i,qs}\big), \\
  i_{i,\mathrm{cool}}^{dq}(t) &= \frac{S_{i}^{\mathrm{cool}}}{S_{\mathrm{base}}}\,i_{i,s}^{dq}(t).
\end{align}
\end{subequations}
The motor therefore has five differential states ($\psi_{i,ds},\psi_{i,qs},\psi_{i,dr},\psi_{i,qr},s_i$), which \eqref{eq:im_v_ds}--\eqref{eq:im_v_qr} and \eqref{eq:im_slip} determine, and five algebraic variables ($i_{i,ds},i_{i,qs},i_{i,dr},i_{i,qr},T_{i,e}$), which \eqref{eq:im_f_ds}--\eqref{eq:im_f_qr} and \eqref{eq:im_te} determine. These ten equations close the motor model for a given LV-bus voltage \eqref{eq:im_vs}; \eqref{eq:im_out} only evaluates outputs.

\subsection{Uninterruptible Power Supply: Double-Conversion UPS}\label{sec:ups}
The double-conversion UPS supplies the IT load through a rectifier, a DC bus, and an output inverter. During normal operation, power flows continuously from the grid through the rectifier and DC bus to the output inverter and IT racks. A battery connects to the DC bus
through a bidirectional DC-to-DC converter and charges or discharges to balance any difference between the rectifier output and the inverter demand. The DC-bus voltage $V_{i,\mathrm{dc}}$ provides the common electrical state that couples the rectifier, output inverter, bidirectional converter, and battery. The capacitor power balance described in Section~\ref{sec:dcbus} governs its dynamics.

\subsubsection{Rectifier Model}
The rectifier uses a grid-following, three-phase, two-level voltage-source converter with $dq$~frame control~\cite{ercot2026dcload}. Its \(d\)-axis regulates the DC-bus voltage, while its \(q\)-axis regulates reactive power. It shares the PLL and inner current-control structure of the transmission-connected GFL inverter but reverses the power sign convention, with positive \(P_{i,\mathrm{rect}}\) denoting AC-to-DC conversion.

\emph{Base quantities:} With $V_{\mathrm{LL,rms}}=480$~V and rectifier rating $S_n$,
\begin{equation}\label{eq:rect_base}
  V_{\mathrm{base}}=\tfrac{\sqrt2}{\sqrt3}V_{\mathrm{LL,rms}}, \quad I_{\mathrm{base}}=\tfrac{\sqrt2 S_n}{\sqrt3 V_{\mathrm{LL,rms}}}, \quad Z_{\mathrm{base}}=\tfrac{V_{\mathrm{base}}}{I_{\mathrm{base}}}, \quad L_{\mathrm{base}}=\tfrac{Z_{\mathrm{base}}}{\omega_0}.
\end{equation}

\subsubsection{Phase-Locked Loop (PLL)}
We model the phase-locked loop (PLL) of rectifier in the data center with index $i\in\mathcal{N}_{\mathrm{rect}}^{\mathrm{dc}}
=\{1,\ldots,N_{\mathrm{rect}}^{\mathrm{dc}}\}$
to estimate the grid-voltage phase angle required for synchronous reference-frame control. The PLL transforms the measured three-phase grid voltages into the rotating $dq$ frame using the estimated angle
$\theta_{i,\mathrm{PLL}}^{r}(t)$ and regulates the quadrature-axis
voltage to zero. We measure $\theta_{i,\mathrm{PLL}}^{r}$ relative to the synchronous frame and project the LV-bus voltage onto the PLL frame:
\begin{subequations}\label{eq:rect_park}
\begin{align}
v_{i,g}^{d,r}(t) &= v_{i,\mathrm{LV}}^{d}(t)\cos\theta_{i,\mathrm{PLL}}^{r}(t)+v_{i,\mathrm{LV}}^{q}(t)\sin\theta_{i,\mathrm{PLL}}^{r}(t), \label{eq:rect_park_d}\\
v_{i,g}^{q,r}(t) &= -v_{i,\mathrm{LV}}^{d}(t)\sin\theta_{i,\mathrm{PLL}}^{r}(t)+v_{i,\mathrm{LV}}^{q}(t)\cos\theta_{i,\mathrm{PLL}}^{r}(t). \label{eq:rect_park_q}
\end{align}
\end{subequations}
We filter the quadrature-axis voltage using a
first-order low-pass filter to attenuate measurement noise:
\begin{subequations}\label{eq:rect_pll}
\begin{align}
\frac{d v_{i,g}^{q,r,\mathrm{fil}}(t)}{dt}
&=
\frac{1}{T_{f,\mathrm{PLL}}^{\mathrm{rect}}}
\left(
v_{i,g}^{q,r}(t)
-
v_{i,g}^{q,r,\mathrm{fil}}(t)
\right), \label{eq:rect_pll_lpf}\\[4pt]
\frac{d I_{i,\mathrm{PLL}}^{r}(t)}{dt}
&=
k_{I,\mathrm{PLL}}^{\mathrm{rect}}\,
v_{i,g}^{q,r,\mathrm{fil}}(t),
\label{eq:rect_pll_integrator}\\[4pt]
\Delta_{i,\mathrm{PLL}}^{r}(t)
&=
k_{P,\mathrm{PLL}}^{\mathrm{rect}}\,
v_{i,g}^{q,r,\mathrm{fil}}(t)
+
I_{i,\mathrm{PLL}}^{r}(t),
\label{eq:rect_pll_pi}\\[4pt]
\frac{d\theta_{i,\mathrm{PLL}}^{r}(t)}{dt}
&=
\omega_0\,\Delta_{i,\mathrm{PLL}}^{r}(t),
\label{eq:rect_pll_angle}\\[4pt]
\hat{\omega}_{i}^{r}(t)
&=
\omega_0+\frac{d\theta_{i,\mathrm{PLL}}^{r}(t)}{dt}
=\omega_0\big(1+\Delta_{i,\mathrm{PLL}}^{r}(t)\big).
\label{eq:rect_pll_omega}
\end{align}
\end{subequations}
The PI controller in \eqref{eq:rect_pll_pi} generates a frequency deviation proportional to the voltage phase error. Integrating this deviation yields the estimated grid angle and synchronizes the rectifier reference frame with the grid; \eqref{eq:rect_pll_omega} gives the estimated angular frequency used in the decoupling terms of the current controller. The PLL therefore has three differential states ($v_{i,g}^{q,r,\mathrm{fil}},I_{i,\mathrm{PLL}}^{r},\theta_{i,\mathrm{PLL}}^{r}$), which \eqref{eq:rect_pll_lpf}, \eqref{eq:rect_pll_integrator}, and \eqref{eq:rect_pll_angle} determine, and four algebraic variables ($v_{i,g}^{d,r},v_{i,g}^{q,r},\Delta_{i,\mathrm{PLL}}^{r},\hat{\omega}_{i}^{r}$), which \eqref{eq:rect_park_d}, \eqref{eq:rect_park_q}, \eqref{eq:rect_pll_pi}, and \eqref{eq:rect_pll_omega} determine, for a given LV-bus voltage $(v_{i,\mathrm{LV}}^{d},v_{i,\mathrm{LV}}^{q})$.

\subsubsection{Power Controller}

We model the outer controller of rectifier index
$i\in\mathcal{N}_{\mathrm{rect}}^{\mathrm{dc}}$ to regulate the DC-bus voltage and reactive-power exchange at the point of common coupling (PCC). Under
the rectifier sign convention, positive active power denotes power transfer from the AC grid to the DC bus. Accordingly, we compute the instantaneous active and reactive power as
\begin{subequations}\label{eq:rect_PQ_dq}
\begin{align}
P_{i,\mathrm{rect}}(t)
&=
-\frac{3}{2}
\left(
v_{i,g}^{d,r}(t)i_{i,g}^{d,r}(t)
+
v_{i,g}^{q,r}(t)i_{i,g}^{q,r}(t)
\right),
\label{eq:rect_P_dq}\\[2mm]
Q_{i,\mathrm{rect}}(t)
&=
\frac{3}{2}
\left(
v_{i,g}^{q,r}(t)i_{i,g}^{d,r}(t)
-
v_{i,g}^{d,r}(t)i_{i,g}^{q,r}(t)
\right).
\label{eq:rect_Q_dq}
\end{align}
\end{subequations}
Assuming lossless AC-DC conversion, the rectifier injects the following current into the DC bus:
\begin{subequations}\label{eq:rect_dc_coupling}
\begin{align}
I_{i,\mathrm{DC}}(t)
&=
\frac{P_{i,\mathrm{rect}}(t)}
     {V_{i,\mathrm{dc}}(t)},\\
I_{i,\mathrm{load}}(t)
&=
\frac{P_{i,\mathrm{load}}(t)}
     {V_{i,\mathrm{dc}}(t)}.
\end{align}
\end{subequations}

We filter the measured DC-bus voltage and reactive power using first-order low-pass filters:
\begin{subequations}\label{eq:rect_outer_fil}
\begin{align}
\frac{dV_{i,\mathrm{dc}}^{\mathrm{fil}}(t)}{dt}
&=
\frac{1}{T_{f,\mathrm{PQ}}^{\mathrm{rect}}}
\left(
V_{i,\mathrm{dc}}(t)
-
V_{i,\mathrm{dc}}^{\mathrm{fil}}(t)
\right), \label{eq:rect_vdc_fil}\\
\frac{dQ_{i,\mathrm{rect}}^{\mathrm{fil}}(t)}{dt}
&=
\frac{1}{T_{f,\mathrm{PQ}}^{\mathrm{rect}}}
\left(
Q_{i,\mathrm{rect}}(t)
-
Q_{i,\mathrm{rect}}^{\mathrm{fil}}(t)
\right). \label{eq:rect_q_fil}
\end{align}
\end{subequations}

\begin{figure}[H]
    \centering
    \includegraphics[width=0.75\columnwidth,keepaspectratio]{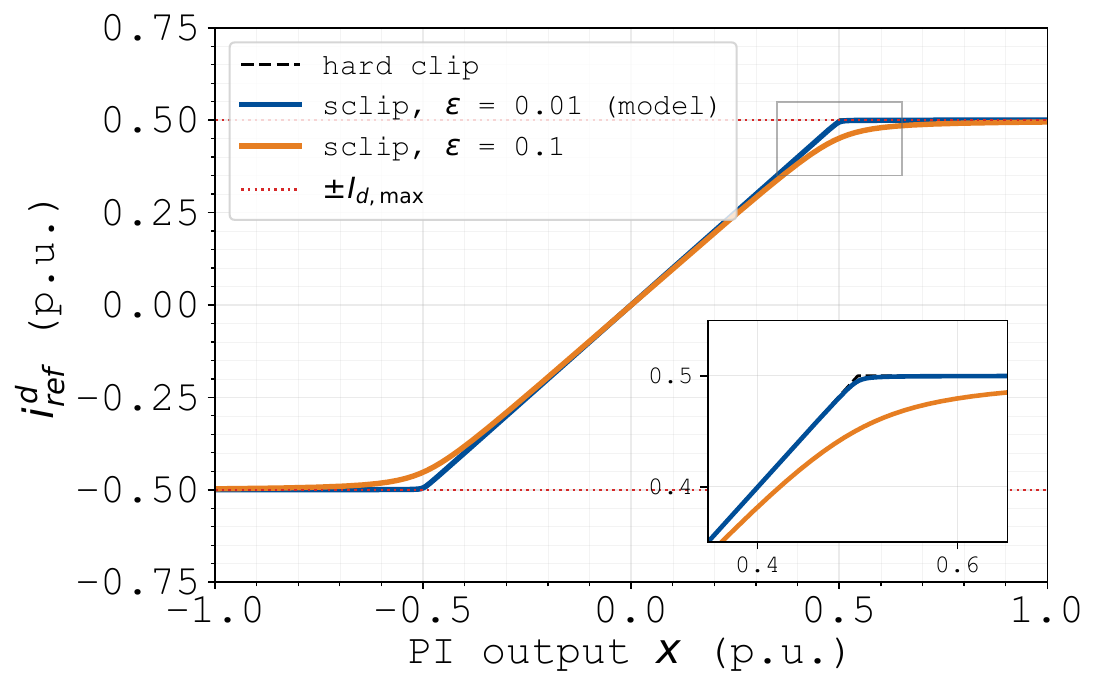}
    \caption{Smooth limiter $\sclip$ for the rectifier current limit $\pm I_{i,d,\max}^{r} = \pm0.5$~p.u. compared with
    the hard clip, for the model's $\varepsilon = 0.01$ and $\varepsilon = 0.1$.}
    \label{fig:sclip}
\end{figure}

The $d$-axis PI controller regulates the DC-bus voltage and generates
the active-current reference. The leading negative sign follows from
the rectifier convention: when
$V_{i,\mathrm{dc}}<V_{i,\mathrm{dc,ref}}$, the controller makes the
$d$-axis current more negative and increases the AC-to-DC power
transfer. We normalize the voltage error by $V_{i,\mathrm{dc,ref}}$ so that the gains act on a per-unit error, and we apply a smooth current limit to represent the rectifier
current capability:
\begin{subequations}\label{eq:rect_daxis_outer}
\begin{align}
\frac{dI_{i,\mathrm{Vdc}}^{d,r}(t)}{dt}
&=
k_{I,\mathrm{Vdc}}^{\mathrm{rect}}
\frac{V_{i,\mathrm{dc,ref}}-V_{i,\mathrm{dc}}^{\mathrm{fil}}(t)}{V_{i,\mathrm{dc,ref}}}
-
\alpha_{\mathrm{leak}}^{\mathrm{rect}}
I_{i,\mathrm{Vdc}}^{d,r}(t), \label{eq:rect_vdc_int}\\
i_{i,g,\mathrm{ref}}^{d,r}(t)
&=
\sclip\!\left(
-\left[
k_{P,\mathrm{Vdc}}^{\mathrm{rect}}
\frac{V_{i,\mathrm{dc,ref}}-V_{i,\mathrm{dc}}^{\mathrm{fil}}(t)}{V_{i,\mathrm{dc,ref}}}
+
I_{i,\mathrm{Vdc}}^{d,r}(t)
\right],
-I_{i,d,\max}^{r},
I_{i,d,\max}^{r}
\right). \label{eq:rect_id_ref}
\end{align}
\end{subequations}
where $I_{i,d,\max}^{r}$ denotes the rectifier $d$-axis current limit. The smooth limiter $\sclip$ of Section~\ref{sec:smooth} rounds the corners of the hard limit over a width of about $\varepsilon$, so that the reference stays continuously differentiable (Fig.~\ref{fig:sclip}).

The $q$-axis PI controller regulates the reactive power and generates the reactive-current reference. We set $Q_{i,\mathrm{ref}}^{r}(t)=0$ for unity power factor operation:
\begin{subequations}\label{eq:rect_qaxis_outer}
\begin{align}
\frac{dI_{i,Q}^{q,r}(t)}{dt}
&=
k_{I,\mathrm{PQ}}^{\mathrm{rect}}
\left(
Q_{i,\mathrm{ref}}^{r}(t)
-
Q_{i,\mathrm{rect}}^{\mathrm{fil}}(t)
\right)
-
\alpha_{\mathrm{leak}}^{\mathrm{rect}}
I_{i,Q}^{q,r}(t), \label{eq:rect_q_int}\\
i_{i,g,\mathrm{ref}}^{q,r}(t)
&=
-\left[
k_{P,\mathrm{PQ}}^{\mathrm{rect}}
\left(
Q_{i,\mathrm{ref}}^{r}(t)
-
Q_{i,\mathrm{rect}}^{\mathrm{fil}}(t)
\right)
+
I_{i,Q}^{q,r}(t)
\right]. \label{eq:rect_iq_ref}
\end{align}
\end{subequations}

\begin{figure}[H]
    \centering
    \includegraphics[width=0.75\columnwidth,keepaspectratio]{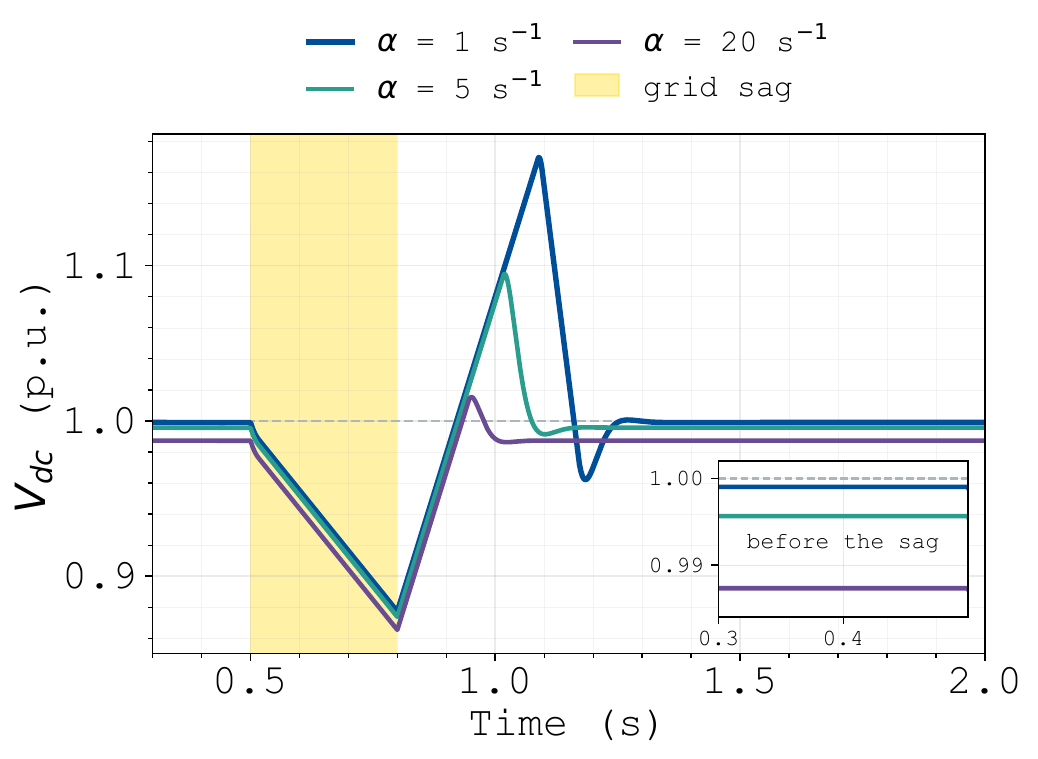}
    \caption{DC-bus voltage of the rectifier loop~\eqref{eq:rect_daxis_outer} for three leakage rates
    $\alpha_{\mathrm{leak}}^{\mathrm{rect}}$, with the tuning of Table~\ref{tab:dc_rect_params}, for an illustrative grid
    sag ($0.6$~p.u. for $0.3$~s, battery disabled) that holds the rectifier at its current limit.}
    \label{fig:leaky}
\end{figure}

The leakage term $\alpha_{\mathrm{leak}}^{\mathrm{rect}}$ gradually reduces the accumulated integrator state, thereby limiting windup when the rectifier reaches its current limit during a voltage sag~\cite{weitenberg2018robust}. After the sag clears, this term allows the controller to leave saturation sooner and restore normal DC-bus voltage regulation more quickly. The price of the leakage is a small steady-state droop: the leaky integrator holds its value only with a constant error, $k_{I,\mathrm{Vdc}}^{\mathrm{rect}}e_{0} = \alpha_{\mathrm{leak}}^{\mathrm{rect}}I_{i,\mathrm{Vdc}}^{d,r}$, so the DC-bus voltage settles at $V_{i,\mathrm{dc,ref}}(1-e_0)$ with $e_0 = |i_{i,g,\mathrm{ref}}^{d,r}|/\big(k_{P,\mathrm{Vdc}}^{\mathrm{rect}} + k_{I,\mathrm{Vdc}}^{\mathrm{rect}}/\alpha_{\mathrm{leak}}^{\mathrm{rect}}\big)$. Fig.~\ref{fig:leaky} illustrates this trade-off for the loop of~\eqref{eq:rect_daxis_outer} with the tuning of Table~\ref{tab:dc_rect_params}: a larger leakage returns the DC-bus voltage faster and with less overshoot after a sag that drives the rectifier to its current limit, but it lowers the voltage before the sag. With $\alpha_{\mathrm{leak}}^{\mathrm{rect}} = 1~\mathrm{s}^{-1}$, the droop is $0.1\%$ ($799.2~\mathrm{V}$).

The power controller therefore has four differential states ($V_{i,\mathrm{dc}}^{\mathrm{fil}},Q_{i,\mathrm{rect}}^{\mathrm{fil}},I_{i,\mathrm{Vdc}}^{d,r},I_{i,Q}^{q,r}$), which \eqref{eq:rect_vdc_fil}, \eqref{eq:rect_q_fil}, \eqref{eq:rect_vdc_int}, and \eqref{eq:rect_q_int} determine, and four algebraic variables ($P_{i,\mathrm{rect}},Q_{i,\mathrm{rect}},i_{i,g,\mathrm{ref}}^{d,r},i_{i,g,\mathrm{ref}}^{q,r}$), which \eqref{eq:rect_P_dq}, \eqref{eq:rect_Q_dq}, \eqref{eq:rect_id_ref}, and \eqref{eq:rect_iq_ref} determine, for given PLL-frame voltages $(v_{i,g}^{d,r},v_{i,g}^{q,r})$, AC currents $(i_{i,g}^{d,r},i_{i,g}^{q,r})$, and DC-bus voltage $V_{i,\mathrm{dc}}$; \eqref{eq:rect_dc_coupling} only evaluates outputs.

\subsubsection{Current Controller}
We model the inner current controller of the rectifier index $i\in\mathcal{N}_{\mathrm{rect}}^{\mathrm{dc}}$ to regulate the AC-side currents in the synchronous reference frame. We filter the measured currents
using first-order low-pass filters with time constant
$T_{f,\mathrm{ig}}^{\mathrm{rect}}$:
\begin{subequations}\label{eq:rect_ig_fil}
\begin{align}
\frac{d i_{i,g}^{d,r,\mathrm{fil}}(t)}{dt}
&=
\frac{1}{T_{f,\mathrm{ig}}^{\mathrm{rect}}}
\left(
i_{i,g}^{d,r}(t)
-
i_{i,g}^{d,r,\mathrm{fil}}(t)
\right), \label{eq:rect_ifil_d}\\
\frac{d i_{i,g}^{q,r,\mathrm{fil}}(t)}{dt}
&=
\frac{1}{T_{f,\mathrm{ig}}^{\mathrm{rect}}}
\left(
i_{i,g}^{q,r}(t)
-
i_{i,g}^{q,r,\mathrm{fil}}(t)
\right). \label{eq:rect_ifil_q}
\end{align}
\end{subequations}

We regulate the filtered currents to track the references
$\big(i_{i,g,\mathrm{ref}}^{d,r}(t),
i_{i,g,\mathrm{ref}}^{q,r}(t)\big)$ using PI controllers with leaky
integrator states:
\begin{subequations}\label{eq:rect_PI_cc}
\begin{align}
\frac{dI_{i,\mathrm{cc}}^{d,r}(t)}{dt}
&=
k_{I,\mathrm{ig}}^{\mathrm{rect}}
\left(
i_{i,g,\mathrm{ref}}^{d,r}(t)
-
i_{i,g}^{d,r,\mathrm{fil}}(t)
\right)
-
\alpha_{\mathrm{leak,cc}}^{\mathrm{rect}}
I_{i,\mathrm{cc}}^{d,r}(t), \label{eq:rect_cc_Id}\\
\epsilon_i^{d,r}(t)
&=
k_{P,\mathrm{ig}}^{\mathrm{rect}}
\left(
i_{i,g,\mathrm{ref}}^{d,r}(t)
-
i_{i,g}^{d,r,\mathrm{fil}}(t)
\right)
+
I_{i,\mathrm{cc}}^{d,r}(t), \label{eq:rect_cc_ed}\\
\frac{dI_{i,\mathrm{cc}}^{q,r}(t)}{dt}
&=
k_{I,\mathrm{ig}}^{\mathrm{rect}}
\left(
i_{i,g,\mathrm{ref}}^{q,r}(t)
-
i_{i,g}^{q,r,\mathrm{fil}}(t)
\right)
-
\alpha_{\mathrm{leak,cc}}^{\mathrm{rect}}
I_{i,\mathrm{cc}}^{q,r}(t), \label{eq:rect_cc_Iq}\\
\epsilon_i^{q,r}(t)
&=
k_{P,\mathrm{ig}}^{\mathrm{rect}}
\left(
i_{i,g,\mathrm{ref}}^{q,r}(t)
-
i_{i,g}^{q,r,\mathrm{fil}}(t)
\right)
+
I_{i,\mathrm{cc}}^{q,r}(t). \label{eq:rect_cc_eq}
\end{align}
\end{subequations}

We generate the rectifier voltage commands by augmenting the PI
outputs with decoupling terms and grid-voltage feedforward:
\begin{subequations}\label{eq:rect_e_ac_dq}
\begin{align}
e_{i,\mathrm{rect}}^{d}(t)
&=
\epsilon_i^{d,r}(t)
-
\hat{\omega}_{i}^{r}(t)L_f^{\mathrm{rect}} i_{i,g}^{q,r}(t)
+
v_{i,g}^{d,r}(t), \label{eq:rect_ed}\\
e_{i,\mathrm{rect}}^{q}(t)
&=
\epsilon_i^{q,r}(t)
+
\hat{\omega}_{i}^{r}(t)L_f^{\mathrm{rect}} i_{i,g}^{d,r}(t)
+
v_{i,g}^{q,r}(t). \label{eq:rect_eq}
\end{align}
\end{subequations}

The rectifier AC currents follow the $dq$-frame dynamics of the $R_f^{\mathrm{rect}}$--$L_f^{\mathrm{rect}}$ filter between the converter voltage and the LV bus:
%
\begin{subequations}\label{eq:rect_KVL}
\begin{align}
L_f^{\mathrm{rect}}\frac{d i_{i,g}^{d,r}(t)}{dt}
&=
-R_f^{\mathrm{rect}} i_{i,g}^{d,r}(t)
+
\hat{\omega}_{i}^{r}(t)L_f^{\mathrm{rect}} i_{i,g}^{q,r}(t)
+
e_{i,\mathrm{rect}}^{d}(t)
-
v_{i,g}^{d,r}(t), \label{eq:rect_kvl_d}\\
L_f^{\mathrm{rect}}\frac{d i_{i,g}^{q,r}(t)}{dt}
&=
-R_f^{\mathrm{rect}} i_{i,g}^{q,r}(t)
-
\hat{\omega}_{i}^{r}(t)L_f^{\mathrm{rect}} i_{i,g}^{d,r}(t)
+
e_{i,\mathrm{rect}}^{q}(t)
-
v_{i,g}^{q,r}(t). \label{eq:rect_kvl_q}
\end{align}
\end{subequations}

The current controller therefore has six differential states ($i_{i,g}^{d,r,\mathrm{fil}},i_{i,g}^{q,r,\mathrm{fil}},I_{i,\mathrm{cc}}^{d,r},I_{i,\mathrm{cc}}^{q,r},i_{i,g}^{d,r},i_{i,g}^{q,r}$), which \eqref{eq:rect_ifil_d}, \eqref{eq:rect_ifil_q}, \eqref{eq:rect_cc_Id}, \eqref{eq:rect_cc_Iq}, \eqref{eq:rect_kvl_d}, and \eqref{eq:rect_kvl_q} determine, and four algebraic variables ($\epsilon_i^{d,r},\epsilon_i^{q,r},e_{i,\mathrm{rect}}^{d},e_{i,\mathrm{rect}}^{q}$), which \eqref{eq:rect_cc_ed}, \eqref{eq:rect_cc_eq}, \eqref{eq:rect_ed}, and \eqref{eq:rect_eq} determine, for given current references $(i_{i,g,\mathrm{ref}}^{d,r},i_{i,g,\mathrm{ref}}^{q,r})$, PLL-frame voltages $(v_{i,g}^{d,r},v_{i,g}^{q,r})$, and estimated frequency $\hat{\omega}_{i}^{r}$.

\begin{remark}\label{rem:afe}
Reactive power remains on the AC side and does not enter the DC-bus current balance in \eqref{eq:rect_dc_coupling}. The $q$-axis controller drives $Q_{i,\mathrm{rect}}^{\mathrm{fil}}(t)$ toward $Q_{i,\mathrm{ref}}^{r}(t)$ by generating $i_{i,g,\mathrm{ref}}^{q,r}(t)$. Setting $Q_{i,\mathrm{ref}}^{r}(t)=0$ enforces unity-power-factor operation, while a nonzero reference enables reactive-power support.
\end{remark}

\subsubsection{Inverter Model}
\label{sec:ups_inv}
We model the UPS output inverter of data center
$i\in\mathcal{N}_{\mathrm{dc}}=\{1,\ldots,N_{\mathrm{dc}}\}$ as a grid-forming converter~\cite{ercot2026dcload}. Unlike the rectifier, the inverter has no PLL: it forms the UPS output bus that supplies the IT load, and it sets its own frequency and voltage magnitude with droop laws on its output active power, its output reactive power, and the DC-bus voltage. The DC bus decouples the two AC sides of the double-conversion UPS, so the UPS output frequency and phase do not follow the grid. In per unit on the unit rating $S_n$,
\begin{subequations}\label{eq:inv_droop}
\begin{align}
  f_{i}^{\mathrm{droop}}(t) &= \Big[1+D_{pf}\big(P_{\mathrm{set}}-P_{i,\mathrm{inv}}(t)\big)\Big]f_{\mathrm{ref}}, \label{eq:inv_f}\\
  V_{i}^{\mathrm{droop}}(t) &= \Big[1+D_{qV}\big(Q_{\mathrm{set}}-Q_{i,\mathrm{inv}}(t)\big)+D_{VV}\Big(\frac{V_{i,\mathrm{dc}}(t)}{V_{i,\mathrm{dc,ref}}}-1\Big)\Big]V_{\mathrm{AC,ord}}, \label{eq:inv_V}
\end{align}
\end{subequations}
where the $P$--$f$ droop shares the load among parallel UPS units without communication, the $Q$--$V$ droop sets the voltage with reactive output, and the $V_{\mathrm{AC}}$--$V_{\mathrm{DC}}$ droop lowers the AC output voltage when the DC bus sags. We integrate the droop frequency to obtain the inverter angle relative to the synchronous frame, and we regulate the output-voltage magnitude with a single PI loop, without an inner current loop:
\begin{subequations}\label{eq:inv_ctrl}
\begin{align}
  \frac{d\theta_{i}^{\mathrm{inv}}(t)}{dt} &= 2\pi\big(f_{i}^{\mathrm{droop}}(t)-f_{\mathrm{ref}}\big), \label{eq:inv_theta}\\
  \frac{dx_{i}^{V}(t)}{dt} &= k_{I,V}^{\mathrm{ups}}\big(V_{i}^{\mathrm{droop}}(t)-V_{i}^{\mathrm{inv}}(t)\big), \label{eq:inv_xV}\\
  v_{i}^{d,\mathrm{inv}}(t) &= k_{P,V}^{\mathrm{ups}}\big(V_{i}^{\mathrm{droop}}(t)-V_{i}^{\mathrm{inv}}(t)\big)+x_{i}^{V}(t), \label{eq:inv_vd}\\
  v_{i}^{q,\mathrm{inv}}(t) &= 0, \label{eq:inv_vq}\\
  V_{i}^{\mathrm{inv}}(t) &= \sqrt{\big(v_{i}^{d,\mathrm{inv}}(t)\big)^2+\big(v_{i}^{q,\mathrm{inv}}(t)\big)^2}. \label{eq:inv_Vmag}
\end{align}
\end{subequations}
We use the average model of the inverter, an ideal controlled voltage source with the output LC filter neglected~\cite{ercot2026dcload}, so $(v_{i}^{d,\mathrm{inv}},v_{i}^{q,\mathrm{inv}})$ is the UPS output-bus voltage in the inverter frame of angle $\theta_{i}^{\mathrm{inv}}$. The constant-power computing load~\eqref{eq:it_ref} draws the current $(i_{i}^{d,\mathrm{inv}},i_{i}^{q,\mathrm{inv}})$ that satisfies
\begin{subequations}\label{eq:inv_load}
\begin{align}
  P_{i}^{\mathrm{IT,u}}(t) &= v_{i}^{d,\mathrm{inv}}(t)\,i_{i}^{d,\mathrm{inv}}(t)+v_{i}^{q,\mathrm{inv}}(t)\,i_{i}^{q,\mathrm{inv}}(t), \label{eq:inv_loadP}\\
  Q_{i}^{\mathrm{IT,u}}(t) &= v_{i}^{q,\mathrm{inv}}(t)\,i_{i}^{d,\mathrm{inv}}(t)-v_{i}^{d,\mathrm{inv}}(t)\,i_{i}^{q,\mathrm{inv}}(t), \label{eq:inv_loadQ}
\end{align}
\end{subequations}
and the inverter delivers
\begin{subequations}\label{eq:inv_PQ}
\begin{align}
  P_{i,\mathrm{inv}}(t) &= v_{i}^{d,\mathrm{inv}}(t)\,i_{i}^{d,\mathrm{inv}}(t)+v_{i}^{q,\mathrm{inv}}(t)\,i_{i}^{q,\mathrm{inv}}(t), \label{eq:inv_P}\\
  Q_{i,\mathrm{inv}}(t) &= v_{i}^{q,\mathrm{inv}}(t)\,i_{i}^{d,\mathrm{inv}}(t)-v_{i}^{d,\mathrm{inv}}(t)\,i_{i}^{q,\mathrm{inv}}(t), \label{eq:inv_Q}
\end{align}
\end{subequations}
whose active part, $P_{i,\mathrm{inv}}^{W}=S_n\,P_{i,\mathrm{inv}}$, leaves the DC bus (Section~\ref{sec:dcbus}). Because the IT load is constant power, $P_{i,\mathrm{inv}}=P_{i}^{\mathrm{IT,u}}$ at every instant; the droop laws shape only the UPS output voltage and frequency. The inverter therefore has two differential states ($\theta_{i}^{\mathrm{inv}},x_{i}^{V}$), which \eqref{eq:inv_theta} and \eqref{eq:inv_xV} determine, and nine algebraic variables ($f_{i}^{\mathrm{droop}},V_{i}^{\mathrm{droop}},v_{i}^{d,\mathrm{inv}},v_{i}^{q,\mathrm{inv}},V_{i}^{\mathrm{inv}},i_{i}^{d,\mathrm{inv}},i_{i}^{q,\mathrm{inv}},P_{i,\mathrm{inv}},Q_{i,\mathrm{inv}}$), which \eqref{eq:inv_f}, \eqref{eq:inv_V}, \eqref{eq:inv_vd}--\eqref{eq:inv_Vmag}, \eqref{eq:inv_loadP}, \eqref{eq:inv_loadQ}, \eqref{eq:inv_P}, and \eqref{eq:inv_Q} determine, for a given IT demand $P_{i}^{\mathrm{IT,u}}(t)$ and DC-bus voltage $V_{i,\mathrm{dc}}$.

Individual server power supplies typically include a bridge rectifier and an LLC converter between the UPS output and the computing load~\cite{ali2026steady}. Rather than model these server-level converters individually, we represent the aggregated servers as a constant-power IT load and determine their demand from a real data center load profile.

\subsubsection{Battery Model}
\begin{figure}[H]
    \centering
    \includegraphics[width=\columnwidth,keepaspectratio]{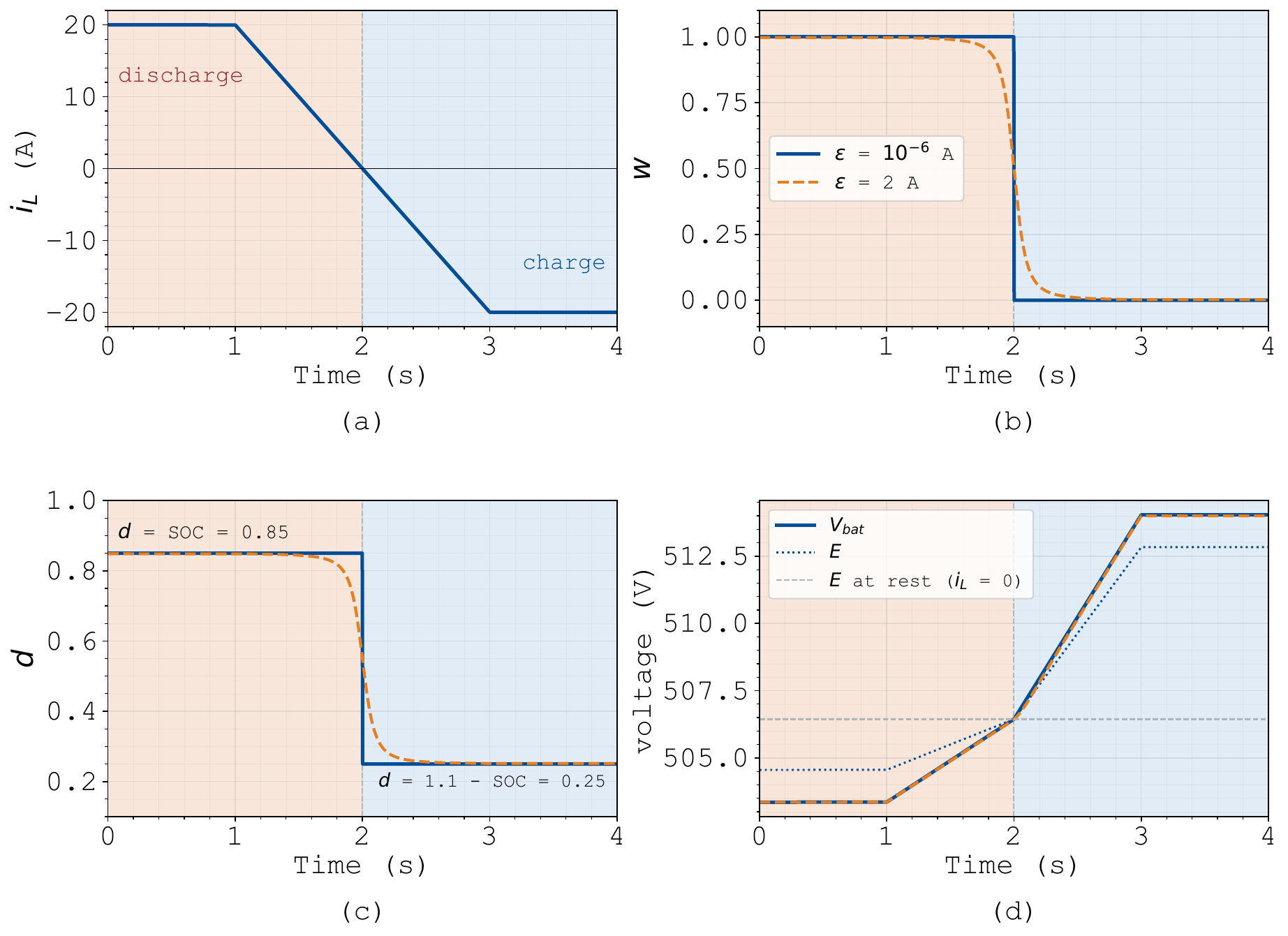}
    \caption{Charge/discharge switch of the battery model at $\mathrm{SOC}_i = 0.85$ for an illustrative current that
    reverses at $t = 2$~s: (a) converter current $i_{i,L}$, (b) weight $w_i$, (c) polarization denominator $d_i$,
    (d) no-load and terminal voltage. Solid: $\varepsilon = 10^{-6}$~A (model).}
    \label{fig:batt_switch}
\end{figure}

We model the UPS battery of data center $i \in \mathcal{N}_{\mathrm{bat}}^{\mathrm{dc}} = \{1, \dots, N_{\mathrm{bat}}^{\mathrm{dc}}\}$ with the modified Shepherd characteristic~\cite{ercot2026dcload}. The bidirectional converter (Section~\ref{sec:bidc}) imposes its inductor current on the battery, $I_{i,\mathrm{bat}} \equiv i_{i,L}$, with $i_{i,L}>0$ for discharge. With $i_{i,L}$ as the input, we compute the state of charge $\mathrm{SOC}_i$, the internal no-load voltage $E_i$, and the terminal voltage $V_{i,\mathrm{bat}}$ as
\begin{subequations}\label{eq:batt_shepherd}
\begin{align}
  \frac{d\,\mathrm{SOC}_i(t)}{dt} &= -\frac{i_{i,L}(t)}{Q_{i,\mathrm{As}}}, \label{eq:batt_shep_soc}\\
  E_i(t) &= E_{i,0} - K_i \frac{Q_i \big(1-\mathrm{SOC}_i(t)\big)}{\mathrm{SOC}_i(t)} - K_i \frac{i_{i,L}(t)}{d_i(t)} + A_i\, e^{-B_i Q_i (1-\mathrm{SOC}_i(t))}, \label{eq:batt_shep_E}\\
  V_{i,\mathrm{bat}}(t) &= E_i(t) - R_{i,\mathrm{int}}\, i_{i,L}(t), \label{eq:batt_shep_vbat}
\end{align}
\end{subequations}
where $Q_{i,\mathrm{As}}=3600\,Q_i$ denotes the capacity in ampere-seconds, $E_{i,0}$ the constant open-circuit voltage, $K_i$ the polarization constant, $Q_i$ the nominal capacity in ampere-hours, $A_i$ and $B_i$ the exponential-zone amplitude and inverse capacity, and $R_{i,\mathrm{int}}$ the internal resistance. The polarization denominator $d_i$ depends on the current direction: it equals $\mathrm{SOC}_i$ during discharge and $1.1-\mathrm{SOC}_i$ during charge. To keep the model continuously differentiable for the NLP solver, we blend the two cases with the smooth sign function of Section~\ref{sec:smooth},
\begin{equation}\label{eq:batt_shep_d}
  d_i(t) = w_i(t)\,\mathrm{SOC}_i(t) + \big(1-w_i(t)\big)\big(1.1-\mathrm{SOC}_i(t)\big),
  \qquad w_i(t) = \tfrac12\Big(1+\ssign\big(i_{i,L}(t)\big)\Big),
\end{equation}
so that $w_i\to1$ (discharge) for $i_{i,L}\gg\varepsilon$, $w_i\to0$ (charge) for $i_{i,L}\ll-\varepsilon$, and $w_i=\tfrac12$ at $i_{i,L}=0$. Fig.~\ref{fig:batt_switch} illustrates this chain for a current that reverses from discharge to charge: the weight $w_i$ switches $d_i$ from $\mathrm{SOC}_i$ to $1.1-\mathrm{SOC}_i$, so the polarization term $-K_i\,i_{i,L}/d_i$ lowers the terminal voltage during discharge and raises it, more steeply, during charge. The voltage stays continuous through the switch because the term vanishes at $i_{i,L}=0$; only its slope $K_i/d_i + R_{i,\mathrm{int}}$ changes. The offset $1.1$ (instead of $1$) keeps $d_i\geq0.1$ at full charge, so that the charging term stays finite~\cite{tremblay2009experimental}.

The battery delivers the power $P_{i,\mathrm{bat}}^{W} = V_{i,\mathrm{bat}}\, i_{i,L}$ to the DC-DC converter. The battery therefore has one differential state ($\mathrm{SOC}_i$), which \eqref{eq:batt_shep_soc} determines, and two algebraic variables ($E_i, V_{i,\mathrm{bat}}$), which \eqref{eq:batt_shep_E} and \eqref{eq:batt_shep_vbat} determine, for a given converter current $i_{i,L}$; \eqref{eq:batt_shep_d} is an explicit function of $\mathrm{SOC}_i$ and $i_{i,L}$ inside~\eqref{eq:batt_shep_E}.

\subsection{Bidirectional DC-DC Converter}\label{sec:bidc}
We model the bidirectional DC-DC converter that connects the battery of data center $i \in \mathcal{N}_{\mathrm{conv}}^{\mathrm{dc}} = \{1, \dots, N_{\mathrm{conv}}^{\mathrm{dc}}\}$ (low-voltage side, $V_{i,\mathrm{bat}}$) to the UPS DC bus (high-voltage side, $V_{i,\mathrm{dc}}$) with a single averaged model and a continuous duty cycle $D_i \in [0,1]$. The ERCOT model~\cite{ercot2026dcload} uses two mode-switched averaged circuits, a buck circuit for charging and a boost circuit for discharging; we replace them with one continuous model so that the NLP remains differentiable when the current changes direction. The inductor current $i_{i,L}$, positive for discharge, obeys
\begin{equation}\label{eq:bidc_iL}
  L_{i,\mathrm{dc}} \frac{d\,i_{i,L}(t)}{dt} = D_i(t)\, V_{i,\mathrm{dc}}(t) - \big(1 - D_i(t)\big) V_{i,\mathrm{bat}}(t) - R_{i,\mathrm{dc}}\, i_{i,L}(t),
\end{equation}
where $L_{i,\mathrm{dc}}$ denotes the inductance and $R_{i,\mathrm{dc}}$ the series resistance. In the ideal steady state ($di_{i,L}/dt=0$, $R_{i,\mathrm{dc}}=0$), \eqref{eq:bidc_iL} gives $V_{i,\mathrm{bat}}/V_{i,\mathrm{dc}}=D_i/(1-D_i)$, that is, $D_i=V_{i,\mathrm{bat}}/(V_{i,\mathrm{dc}}+V_{i,\mathrm{bat}})$. The direction of power flow therefore follows the sign of $i_{i,L}$, not the value of $D_i$: a duty slightly above this value drives discharge, and a duty slightly below it drives charging. The converter exchanges the power $P_{i,\mathrm{bat}}^{W}=V_{i,\mathrm{bat}}\,i_{i,L}$ with the DC bus~\eqref{eq:dcbus}.

The battery supports the DC bus only when the bus voltage falls below a dead band $(1-\delta_{\mathrm{db}})V_{i,\mathrm{dc,ref}}$, so that the rectifier alone regulates the bus in normal operation. We define the smooth one-sided voltage error, a leaky integrator, a charger current that refills the battery toward $\mathrm{SOC}_{i,0}$ while the bus is healthy, and the battery-current reference:
\begin{subequations}\label{eq:ups_vdc_loop}
\begin{align}
  \tilde{e}_{i}(t) &= \smax\big((1-\delta_{\mathrm{db}})V_{i,\mathrm{dc,ref}} - V_{i,\mathrm{dc}}(t),\, 0\big), \label{eq:bidc_err}\\
  \frac{d\,x_{i,\mathrm{Vdc}}(t)}{dt} &= k_{I,\mathrm{Vdc}}^{\mathrm{bi}}\, \tilde{e}_{i}(t) - \alpha^{\mathrm{bi}}\, x_{i,\mathrm{Vdc}}(t), \label{eq:bidc_x}\\
  I_{i,\mathrm{ch}}(t) &= I_{\mathrm{ch,max}}\, \phi_i(t)\, \gamma_i(t), \label{eq:bidc_ich}\\
  I_{i,\mathrm{ref}}^{\mathrm{bat}}(t) &= k_{P,\mathrm{Vdc}}^{\mathrm{bi}}\, \tilde{e}_{i}(t) + x_{i,\mathrm{Vdc}}(t) - I_{i,\mathrm{ch}}(t), \label{eq:bidc_iref}
\end{align}
\end{subequations}
where the SOC taper $\phi_i$ and the bus-health gate $\gamma_i$ are
\begin{equation}\label{eq:bidc_taper}
  \phi_i = \frac{\sclip\big((\mathrm{SOC}_{i,0}-\mathrm{SOC}_i)/\Delta_{\mathrm{SOC}},\,0,\,1\big)-\phi_0}{1-\phi_0}, \qquad
  \gamma_i = 1-\sclip\big(\tilde{e}_{i}/(0.01\,V_{i,\mathrm{dc,ref}}),\,0,\,1\big),
\end{equation}
with $\phi_0=\sclip(0,0,1)$, so that the charger is exactly off at $\mathrm{SOC}_i=\mathrm{SOC}_{i,0}$ and switches off as soon as the bus sags. A sag ($\tilde{e}_i>0$) therefore commands discharge ($I_{i,\mathrm{ref}}^{\mathrm{bat}}>0$); the leakage $\alpha^{\mathrm{bi}}$ returns the integrator to zero once the bus recovers. The inner loop combines the feedforward duty with proportional current control and a smooth duty limit:
\begin{subequations}\label{eq:bidc_duty}
\begin{align}
  e_{i,I}(t) &= I_{i,\mathrm{ref}}^{\mathrm{bat}}(t) - i_{i,L}(t), \label{eq:bidc_eI}\\
  D_i(t) &= \sclip\Big(\frac{V_{i,\mathrm{bat}}(t)}{V_{i,\mathrm{dc}}(t) + V_{i,\mathrm{bat}}(t)} + k_{P,\mathrm{ig}}^{\mathrm{bi}}\, e_{i,I}(t),\, D_{\min},\, D_{\max}\Big). \label{eq:bidc_D}
\end{align}
\end{subequations}
Within the duty limits, substituting~\eqref{eq:bidc_D} into~\eqref{eq:bidc_iL} gives $L_{i,\mathrm{dc}}\,di_{i,L}/dt = k_{P,\mathrm{ig}}^{\mathrm{bi}}(V_{i,\mathrm{dc}}+V_{i,\mathrm{bat}})\,e_{i,I} - R_{i,\mathrm{dc}}\,i_{i,L}$, a first-order current loop with time constant $L_{i,\mathrm{dc}}/\big(k_{P,\mathrm{ig}}^{\mathrm{bi}}(V_{i,\mathrm{dc}}+V_{i,\mathrm{bat}})\big)\approx0.1$~ms. The converter therefore has two differential states ($i_{i,L},x_{i,\mathrm{Vdc}}$), which \eqref{eq:bidc_iL} and \eqref{eq:bidc_x} determine, and five algebraic variables ($\tilde{e}_{i},I_{i,\mathrm{ch}},I_{i,\mathrm{ref}}^{\mathrm{bat}},e_{i,I},D_i$), which \eqref{eq:bidc_err}, \eqref{eq:bidc_ich}, \eqref{eq:bidc_iref}, \eqref{eq:bidc_eI}, and \eqref{eq:bidc_D} determine, for a given DC-bus voltage $V_{i,\mathrm{dc}}$, battery voltage $V_{i,\mathrm{bat}}$, and state of charge $\mathrm{SOC}_i$; \eqref{eq:bidc_taper} defines explicit functions inside~\eqref{eq:bidc_ich}. The dead band, the leakage, and the charger are additions of this study.

\subsection{DC-Bus Power Balance and Dynamic Coupling}\label{sec:dcbus}
We formulate the instantaneous power balance across the DC-link capacitor of data center $i \in \mathcal{N}_{\mathrm{dc}} = \{1, \dots, N_{\mathrm{dc}}\}$, which couples the rectifier, the battery converter, and the output inverter:
\begin{equation}\label{eq:dcbus}
  C_{i,\mathrm{dc}} \frac{d\,V_{i,\mathrm{dc}}(t)}{dt} = \frac{P_{i,\mathrm{rect}}^{W}(t) + V_{i,\mathrm{bat}}(t)\, i_{i,L}(t) - P_{i,\mathrm{inv}}^{W}(t)}{V_{i,\mathrm{dc}}(t)},
\end{equation}
where $C_{i,\mathrm{dc}}$ denotes the DC-link capacitance, $P_{i,\mathrm{rect}}^{W}=S_n P_{i,\mathrm{rect}}$ the power the rectifier delivers to the bus, $V_{i,\mathrm{bat}}\, i_{i,L}$ the power the battery converter injects ($i_{i,L}>0$ for discharge), and $P_{i,\mathrm{inv}}^{W}=S_n P_{i,\mathrm{inv}}$ the power the output inverter draws to serve the IT load. The DC bus therefore has one differential state ($V_{i,\mathrm{dc}}$), which \eqref{eq:dcbus} determines, for given rectifier, battery, and inverter powers.

\section{Composite Data-Center Load and Network Interface}\label{sec:composite}
We aggregate the three low-voltage (LV) branches into the total active and reactive power drawn by data center $i \in \mathcal{N}_{\mathrm{dc}} = \{1, \dots, N_{\mathrm{dc}}\}$ at its point of interconnection (POI). The total demand comprises the UPS-buffered IT server load, the facility cooling infrastructure, and the remaining auxiliary load. For the IT load, we represent the demand either via the voltage-dependent variable-resistance characteristic in~\eqref{eq:it_load} or through the detailed UPS model established in Section~\ref{sec:ups}. For the cooling system, we compute the cooling load using the induction motor model~\ref{sec:cooling}. Finally, we model the residual auxiliary demand as a constant-impedance characteristic:
\begin{equation}\label{eq:composite_z}
  P_{i}^{z} = G_{i}^{z} (V_{i}^{\mathrm{dc}})^2, \qquad Q_{i}^{z} = B_{i}^{z} (V_{i}^{\mathrm{dc}})^2,
\end{equation}
where $G_{i}^{z}$ and $B_{i}^{z}$ denote the equivalent shunt conductance and susceptance, and $V_{i}^{\mathrm{dc}}$ represents the LV bus voltage magnitude. Summing the respective branch components yields the net active and reactive power demand at the LV bus of data center~$i$:
\begin{subequations}\label{eq:composite_total}
\begin{align}
  P_{i}^{\mathrm{LV}} &= P_{i}^{\mathrm{IT}} + P_{i}^{\mathrm{cool}} + P_{i}^{z}, \label{eq:composite_P}\\
  Q_{i}^{\mathrm{LV}} &= Q_{i}^{\mathrm{IT}} + Q_{i}^{\mathrm{cool}} + Q_{i}^{z}. \label{eq:composite_Q}
\end{align}
\end{subequations}
The transformer cascade of Section~\ref{sec:xfmr} steps down the transmission-bus voltage to drive the LV-bus voltage $(v_{i,\mathrm{LV}}^{d}, v_{i,\mathrm{LV}}^{q})$ through the current balance~\eqref{eq:xfmr_kcl}. We map the POI demand $(P_{i}^{\mathrm{dc}}, Q_{i}^{\mathrm{dc}})$ of~\eqref{eq:xfmr_out}, which equals $(P_{i}^{\mathrm{LV}}, Q_{i}^{\mathrm{LV}})$ plus the transformer losses, into transmission bus $j$ as the nodal load $(P_j^d, Q_j^d)$.

\subsection{Aggregation of the UPS Units}\label{sec:aggregation}
We represent the IT load of data center $i$ by $N_i^{\mathrm{ups}}$ identical double-conversion UPS units and model one representative unit of rating $S_n$, with the equations of Section~\ref{sec:ups} in per unit on $S_n$. Each unit serves the demand $P_{i,\mathrm{load}}$ at $g_i = 1$, and the fleet scale
\begin{equation}\label{eq:k_ups}
  k_i^{\mathrm{ups}} = \frac{N_i^{\mathrm{ups}}\,S_n}{S_{\mathrm{base}}}, \qquad N_i^{\mathrm{ups}} \approx \frac{P_{i,0}^{\mathrm{IT}}}{P_{i,\mathrm{load}}},
\end{equation}
converts the unit to the data center on the system base: the rectifier current that enters the LV-bus balance~\eqref{eq:xfmr_kcl} and the IT power are the unit's values multiplied by $k_i^{\mathrm{ups}}$, $P_i^{\mathrm{IT}} = k_i^{\mathrm{ups}} P_{i,\mathrm{rect}}$. We solve $k_i^{\mathrm{ups}}$, together with $B_i^{z}$, from the steady state at $t = 0$, so that the LV bus absorbs exactly the power-flow load; the solved value includes the converter losses and remains constant during the simulation. Because every unit parameter is expressed in per unit on $S_n$, the fleet behaves as one converter of rating $N_i^{\mathrm{ups}} S_n$ with the same per-unit parameters, while the DC-bus voltage, the battery quantities, and the inverter power keep their physical values of one unit.
\emph{Assumption:} all $N_i^{\mathrm{ups}}$ units of a data center are identical, share the same operating point, and respond in lockstep. The model therefore does not capture differences between units, such as units that reach their current limit or trip at different instants, different states of charge, or staggered reconnection after a disturbance.

\section{Smooth Operators and Discretization}\label{sec:smooth}
Every non-smooth operation (saturation, absolute value, sign, min/max) is replaced by a $\sqrt{\cdot}$-based smooth surrogate so that the resulting NLP is continuously differentiable:
\begin{subequations}\label{eq:smooth_ops}
\begin{align}
  \smax(a,b) &= \tfrac12\big(a+b+\sqrt{(a-b)^2+\varepsilon^2}\big), \\
  \smin(a,b) &= \tfrac12\big(a+b-\sqrt{(a-b)^2+\varepsilon^2}\big), \\
  \sclip(x, \underline{x}, \bar{x}) &= \smin\big(\smax(x, \underline{x}),\, \bar{x}\big), \\
  \ssign(x) &= \dfrac{x}{\sqrt{x^2+\varepsilon^2}}, \qquad |x|_\varepsilon = \sqrt{x^2+\varepsilon^2}.
\end{align}
\end{subequations}
The smoothing width is $\varepsilon = 10^{-2}$ for $\smax$, $\smin$, and $\sclip$ (in the units of the argument) and $\varepsilon = 10^{-6}$~A for $\ssign$. Away from a corner the surrogates equal the exact operation to within $\varepsilon^2/(4|a-b|)$; the largest deviation, $\varepsilon/2$, occurs at the corner itself. Fig.~\ref{fig:sclip} shows $\sclip$ for the rectifier current limit.

Every differential state $x$ with $\dot x = f(x,u)$ is discretized by using the trapezoidal rule with a step $h$,
\begin{equation}\label{eq:trap}
  x_k = x_{k-1} + \frac{h}{2}\big(f_k+f_{k-1}\big),
\end{equation}

\section{Parameters}\label{sec:parameters}

This section collects the parameter values of every model in this appendix. All
symbols follow the conventions of Section~\ref{sec:nomenclature}. Per-unit
quantities use a $S_{\mathrm{base}}=100$~MVA system base unless a device base is
stated explicitly, and $f_0=60$~Hz throughout.

\subsection{Transmission Network and Synchronous Generator}

The network and machine data (bus admittances $G_{ik},B_{ik}$, and the
per-generator $E_{i}'$, $X_{i}^{d'}$, $H_i^{g}$, $D_i^{g}$, $T_{i}^{\mathrm{gov}}$,
$R_i$) are taken from the test-case data set and are therefore not reproduced
here. Table~\ref{tab:net_params} lists the system-wide constants.

\begin{table}[H]
\centering
\caption{Transmission-network constants}
\label{tab:net_params}
\begin{tabularx}{\linewidth}{@{}llXr@{}}
\toprule
Symbol & Description & & Value \\
\midrule
$f_0$              & nominal frequency                & & $60$~Hz \\
$\omega_0=2\pi f_0$& nominal angular frequency        & & $376.99$~rad/s \\
$S_{\mathrm{base}}$& system power base                & & $100$~MVA \\
\bottomrule
\end{tabularx}
\end{table}

\subsection{Grid-Following Inverter}

Table~\ref{tab:gfl_params} gives the inverter filter and controller data. The
per-unit filter values and the current-loop gains depend on the device base, so
both the design expression and its value at the base used in this work\footnote{The value changes depend on the base value of the inter-connected bus}
($V_{\mathrm{LL,rms}}=18.44$~kV, $S_n=310$~MVA) are listed. The current-loop
gains follow internal-model control, which cancels the filter pole and leaves a
first-order loop of bandwidth $1/T_{f,\mathrm{ig}}^{\mathrm{inv}}$.

\begin{table}[H]
\centering
\caption{Grid-following inverter parameters}
\label{tab:gfl_params}
\begin{tabularx}{\linewidth}{@{}lXlr@{}}
\toprule
Symbol & Description & Expression & Value \\
\midrule
\multicolumn{4}{@{}l}{\emph{Per-unit bases}}\\
$V_{\mathrm{base}}$ & voltage base & $\tfrac{\sqrt2}{\sqrt3}V_{\mathrm{LL,rms}}$ & $15.06$~kV \\
$I_{\mathrm{base}}$ & current base & $\tfrac{\sqrt2 S_n}{\sqrt3 V_{\mathrm{LL,rms}}}$ & $13.73$~kA \\
$Z_{\mathrm{base}}$ & impedance base & $V_{\mathrm{base}}/I_{\mathrm{base}}$ & $1.097\ \Omega$ \\
$L_{\mathrm{base}}$ & inductance base & $Z_{\mathrm{base}}/\omega_0$ & $2.910$~mH \\
\addlinespace
\multicolumn{4}{@{}l}{\emph{Output filter}}\\
$R_f^{\mathrm{inv}}$ & filter resistance & --- & $1.73$~m$\Omega$ ($1.577\times10^{-3}$~pu) \\
$L_f^{\mathrm{inv}}$ & filter inductance & --- & $0.459$~mH ($0.1578$~pu) \\
\addlinespace
\multicolumn{4}{@{}l}{\emph{Phase-locked loop}}\\
$T_{f,\mathrm{PLL}}^{\mathrm{inv}}$ & voltage-filter time constant & --- & $1$~ms \\
$k_{P,\mathrm{PLL}}^{\mathrm{inv}}$ & proportional gain & $1/(2\pi f_0 T_{f,\mathrm{PLL}}^{\mathrm{inv}})$ & $2.653$ \\
$k_{I,\mathrm{PLL}}^{\mathrm{inv}}$ & integral gain & $k_{P,\mathrm{PLL}}^{\mathrm{inv}}/(4T_{f,\mathrm{PLL}}^{\mathrm{inv}})$ & $663.1$~s$^{-1}$ \\
\addlinespace
\multicolumn{4}{@{}l}{\emph{Power controller}}\\
$T_{f,\mathrm{PQ}}^{\mathrm{inv}}$ & power-filter time constant & --- & $1$~ms \\
$k_{P,\mathrm{PQ}}^{\mathrm{inv}}$ & proportional gain & --- & $0.5$ \\
$k_{I,\mathrm{PQ}}^{\mathrm{inv}}$ & integral gain & $k_{P,\mathrm{PQ}}^{\mathrm{inv}}/(10T_{f,\mathrm{PQ}}^{\mathrm{inv}})$ & $50$~s$^{-1}$ \\
\addlinespace
\multicolumn{4}{@{}l}{\emph{Current controller}}\\
$T_{f,\mathrm{ig}}^{\mathrm{inv}}$ & current-filter time constant & --- & $1$~ms \\
$k_{P,\mathrm{ig}}^{\mathrm{inv}}$ & proportional gain & $L_{f,\mathrm{pu}}^{\mathrm{inv}}/(2T_{f,\mathrm{ig}}^{\mathrm{inv}})$ & $78.88$ \\
$k_{I,\mathrm{ig}}^{\mathrm{inv}}$ & integral gain & $R_{f,\mathrm{pu}}^{\mathrm{inv}}/(2T_{f,\mathrm{ig}}^{\mathrm{inv}})$ & $0.7886$~s$^{-1}$ \\
\bottomrule
\end{tabularx}
\end{table}

\subsection{Data-Center Transformer Cascade and Load Composition}

Table~\ref{tab:dc_xfmr_params} lists the two step-down transformers and the
split of the data-center demand across its three low-voltage branches. The
leakage reactances are referred from a nameplate rating of $250$~MVA to the
$100$~MVA system base.

\begin{table}[H]
\centering
\caption{Data-center transformer cascade and load composition}
\label{tab:dc_xfmr_params}
\begin{tabularx}{\linewidth}{@{}lXr@{}}
\toprule
Symbol & Description & Value \\
\midrule
$R_{i,1},\ X_{i,1}$ & T1 (HV/MV) leakage resistance and reactance & $0.0020$, $0.020$~pu \\
$R_{i,2},\ X_{i,2}$ & T2 (MV/LV) leakage resistance and reactance & $0.0024$, $0.024$~pu \\
--- & HV (transmission) nominal voltage & $230$~kV \\
--- & MV bus nominal voltage & $34.5$~kV \\
--- & LV bus nominal voltage & $480$~V \\
\addlinespace
--- & IT-load share of data-center demand & $0.60$ \\
--- & cooling-load share of data-center demand & $0.30$ \\
--- & constant-impedance share of data-center demand & $0.10$ \\
\bottomrule
\end{tabularx}
\end{table}

\subsection{Data-Center IT and Miscellaneous Loads}

\begin{table}[H]
\centering
\caption{IT-load and miscellaneous-load parameters}
\label{tab:dc_it_params}
\begin{tabularx}{\linewidth}{@{}lXr@{}}
\toprule
Symbol & Description & Value \\
\midrule
$T_{f,\mathrm{PFC}}$ & PFC voltage-filter time constant, $1/(2\pi\cdot 15~\mathrm{Hz})$ & $10.61$~ms \\
$P_{i,0}^{\mathrm{IT}}$ & initial IT active power ($t=0$) & $0.60\,P_{i}^{\mathrm{LV}}(0)$ \\
$g_{i}(t)$ & normalized workload profile & $g_i(0)=1.0$ \\
$G_{i}^{z}$ & equivalent shunt conductance & $P_{i}^{z,0}/(V_{i}^{\mathrm{dc},0})^{2}$ \\
$B_{i}^{z}$ & equivalent shunt susceptance & $Q_{i}^{z,0}/(V_{i}^{\mathrm{dc},0})^{2}$ \\
\bottomrule
\end{tabularx}
\end{table}

Here $P_{i}^{\mathrm{dc},0}$ and $V_{i}^{\mathrm{dc},0}$ denote the pre-disturbance
data-center demand and LV-bus voltage, so the conductance and susceptance
reproduce the residual branch power at the initial operating point.

\subsection{Data-Center Cooling Load (Induction Motor)}

Table~\ref{tab:dc_motor_params} gives the aggregate cooling-motor data in
per-unit on the cooling-branch rating.

\begin{table}[H]
\centering
\caption{Aggregate cooling-motor parameters}
\label{tab:dc_motor_params}
\begin{tabularx}{\linewidth}{@{}lXr@{}}
\toprule
Symbol & Description & Value \\
\midrule
$R_s$ & stator resistance & $0.03$~pu \\
$X_s=\omega_0L_{ls}$ & stator leakage reactance & $0.10$~pu \\
$X_m=\omega_0L_m$ & magnetizing reactance & $3.0$~pu \\
$R_r$ & rotor resistance & $0.02$~pu \\
$X_r=\omega_0L_{lr}$ & rotor leakage reactance & $0.10$~pu \\
$H^{m}$ & motor-plus-load inertia constant & $0.7$~s \\
$s_0$ & initial slip & $0.02$ \\
$T_{i,L}$ & mechanical load torque & $T_{i,e}|_{s_i=s_0}$ \\
$S_{i}^{\mathrm{cool}}$ & cooling-branch rating & motor draws $0.30\,P_{i}^{\mathrm{LV}}(0)$ at $s_0$ \\
\bottomrule
\end{tabularx}
\end{table}

\subsection{Data-Center UPS Rectifier}

Table~\ref{tab:dc_rect_params} gives the rectifier data on its own $V_{\mathrm{LL,rms}}=480$~V, $S_n=150$~kVA base. The DC-bus-voltage loop is tuned for a bandwidth $f_{\mathrm{bw}}$ and damping $\zeta$ well below the current-loop bandwidth, giving $k_{I,\mathrm{Vdc}}^{\mathrm{rect}}=\omega_n^{2}/K_{\mathrm{plant}}$ and
$k_{P,\mathrm{Vdc}}^{\mathrm{rect}}=2\zeta\omega_n/K_{\mathrm{plant}}$ with
$\omega_n=2\pi f_{\mathrm{bw}}$ and
$K_{\mathrm{plant}}=\tfrac{3}{2}S_n/(C_{i,\mathrm{dc}}V_{i,\mathrm{dc,ref}}^{2})$~\cite{554176}.

\begin{table}[H]
\centering
\caption{UPS rectifier parameters}
\label{tab:dc_rect_params}
\begin{tabularx}{\linewidth}{@{}lXlr@{}}
\toprule
Symbol & Description & Expression & Value \\
\midrule
\multicolumn{4}{@{}l}{\emph{Ratings and per-unit bases}}\\
$V_{\mathrm{LL,rms}}$ & rated line-to-line RMS voltage & --- & $480$~V \\
$S_n$ & rated apparent power & --- & $150$~kVA \\
$V_{\mathrm{base}}$ & voltage base & $\tfrac{\sqrt2}{\sqrt3}V_{\mathrm{LL,rms}}$ & $391.9$~V \\
$I_{\mathrm{base}}$ & current base & $\tfrac{\sqrt2 S_n}{\sqrt3 V_{\mathrm{LL,rms}}}$ & $255.2$~A \\
$Z_{\mathrm{base}}$ & impedance base & $V_{\mathrm{base}}/I_{\mathrm{base}}$ & $1.536\ \Omega$ \\
$L_{\mathrm{base}}$ & inductance base & $Z_{\mathrm{base}}/\omega_0$ & $4.074$~mH \\
\addlinespace
\multicolumn{4}{@{}l}{\emph{Converter reactor}}\\
$R_f^{\mathrm{rect}}$ & filter resistance & --- & $1.73$~m$\Omega$ ($1.126\times10^{-3}$~pu) \\
$L_f^{\mathrm{rect}}$ & filter inductance & --- & $0.459$~mH ($0.1127$~pu) \\
\addlinespace
\multicolumn{4}{@{}l}{\emph{Phase-locked loop}}\\
$T_{f,\mathrm{PLL}}^{\mathrm{rect}}$ & voltage-filter time constant & --- & $1$~ms \\
$k_{P,\mathrm{PLL}}^{\mathrm{rect}}$ & proportional gain & $1/(2\pi f_0 T_{f,\mathrm{PLL}}^{\mathrm{rect}})$ & $2.653$ \\
$k_{I,\mathrm{PLL}}^{\mathrm{rect}}$ & integral gain & $k_{P,\mathrm{PLL}}^{\mathrm{rect}}/(4T_{f,\mathrm{PLL}}^{\mathrm{rect}})$ & $663.1$~s$^{-1}$ \\
\addlinespace
\multicolumn{4}{@{}l}{\emph{Outer loops}}\\
$T_{f,\mathrm{PQ}}^{\mathrm{rect}}$ & outer-loop filter time constant & --- & $1$~ms \\
$f_{\mathrm{bw}},\ \zeta$ & DC-bus-loop bandwidth and damping & --- & $8$~Hz, $0.707$ \\
$k_{P,\mathrm{PQ}}^{\mathrm{rect}}$ & reactive-loop proportional gain & --- & $0.5$ \\
$k_{I,\mathrm{PQ}}^{\mathrm{rect}}$ & reactive-loop integral gain & $k_{P,\mathrm{PQ}}^{\mathrm{rect}}/(10T_{f,\mathrm{PQ}}^{\mathrm{rect}})$ & $50$~s$^{-1}$ \\
$Q_{i,\mathrm{ref}}^{r}$ & reactive-power reference & --- & $0$ (unity p.f.) \\
$I_{i,d,\max}^{r}$ & $d$-axis current limit & --- & $0.50$~pu \\
$\alpha_{\mathrm{leak}}^{\mathrm{rect}}$ & outer-loop anti-windup leakage & --- & $1.0$~s$^{-1}$ \\
\addlinespace
\multicolumn{4}{@{}l}{\emph{Current controller}}\\
$T_{f,\mathrm{ig}}^{\mathrm{rect}}$ & current-filter time constant & --- & $10$~ms \\
$k_{P,\mathrm{ig}}^{\mathrm{rect}}$ & proportional gain & $L_{f,\mathrm{pu}}^{\mathrm{rect}}/(2T_{f,\mathrm{ig}}^{\mathrm{rect}})$ & $5.633$ \\
$k_{I,\mathrm{ig}}^{\mathrm{rect}}$ & integral gain & $R_{f,\mathrm{pu}}^{\mathrm{rect}}/(2T_{f,\mathrm{ig}}^{\mathrm{rect}})$ & $0.0563$~s$^{-1}$ \\
$\alpha_{\mathrm{leak,cc}}^{\mathrm{rect}}$ & current-loop anti-windup leakage & --- & $0.5$~s$^{-1}$ \\
\bottomrule
\end{tabularx}
\end{table}

\subsection{Data-Center UPS Battery}

Table~\ref{tab:dc_batt_params} gives the Shepherd battery data. The exponential
and polarization terms act on the extracted charge $Q_i(1-\mathrm{SOC}_i)$, so
$K_i$ and $B_i$ are expressed per ampere-hour.

\begin{table}[H]
\centering
\caption{UPS battery (Shepherd) parameters}
\label{tab:dc_batt_params}
\begin{tabularx}{\linewidth}{@{}lXr@{}}
\toprule
Symbol & Description & Value \\
\midrule
$Q_i$ & nominal capacity & $120$~Ah \\
$Q_{i,\mathrm{As}}=3600\,Q_i$ & capacity in ampere-seconds & $4.32\times10^{5}$~A\,s \\
$E_{i,0}$ & constant open-circuit voltage & $500$~V \\
$K_i$ & polarization constant & $0.08$~V/Ah \\
$A_i$ & exponential-zone amplitude & $20$~V \\
$B_i$ & exponential-zone inverse capacity & $0.05$~(Ah)$^{-1}$ \\
$R_{i,\mathrm{int}}$ & internal ohmic resistance & $0.060\ \Omega$ \\
$\mathrm{SOC}_{i,0}$ & initial state of charge & $0.85$ \\
$\mathrm{SOC}_{i,\min}$ & minimum state of charge & $0.20$ \\
\bottomrule
\end{tabularx}
\end{table}

\subsection{Data-Center Bidirectional DC--DC Converter and DC Bus}

\begin{table}[H]
\centering
\caption{Bidirectional DC--DC converter and DC-bus parameters}
\label{tab:dc_bidc_params}
\begin{tabularx}{\linewidth}{@{}lXr@{}}
\toprule
Symbol & Description & Value \\
\midrule
$L_{i,\mathrm{dc}}$ & converter filter inductance & $2.5$~mH \\
$R_{i,\mathrm{dc}}$ & converter series resistance & $20$~m$\Omega$ \\
$D_{i,\min},\ D_{i,\max}$ & duty-cycle limits & $0$, $1$ \\
$k_{P,\mathrm{Vdc}}^{\mathrm{bi}}$ & DC-bus-loop proportional gain & $1.5$~A/V \\
$k_{I,\mathrm{Vdc}}^{\mathrm{bi}}$ & DC-bus-loop integral gain & $15.0$~A/(V\,s) \\
$k_{P,\mathrm{ig}}^{\mathrm{bi}}$ & inner current-loop proportional gain & $0.02$ \\
$\delta_{\mathrm{db}}$ & DC-bus dead band before battery support (this study) & $0.02$ \\
$\alpha^{\mathrm{bi}}$ & outer-loop integrator leakage (this study) & $1.0$~s$^{-1}$ \\
$I_{\mathrm{ch,max}}$ & maximum charger current, C/10 (this study) & $12$~A \\
$\Delta_{\mathrm{SOC}}$ & charger taper width below $\mathrm{SOC}_{i,0}$ (this study) & $2\times10^{-5}$ \\
\addlinespace
$C_{i,\mathrm{dc}}$ & DC-link capacitance & $50$~mF \\
$V_{i,\mathrm{dc,ref}}$ & DC-bus voltage reference & $800$~V \\
$S_{\mathrm{inv}}$ & output-inverter rating & $150$~kVA \\
$P_{i,\mathrm{load}}$ & IT-load demand at the inverter output & $80$~kW \\
\bottomrule
\end{tabularx}
\end{table}

\subsection{Data-Center UPS Output Inverter (Grid-Forming)}

Table~\ref{tab:dc_inv_params} gives the droop and voltage-loop data of the grid-forming UPS output inverter, in per unit on the unit rating $S_n=S_{\mathrm{inv}}=150$~kVA. The droop settings follow the ERCOT model defaults~\cite{ercot2026dcload}, whose ``damping'' values divide the power and voltage errors, so $D_{pf}=D_{qV}=1/1000$ and $D_{VV}=1/0.5$. The manual does not list the voltage-loop gains; we choose them so that the closed loop of~\eqref{eq:inv_xV}--\eqref{eq:inv_Vmag} has the time constant $(1+k_{P,V}^{\mathrm{ups}})/k_{I,V}^{\mathrm{ups}}=20$~ms.

\begin{table}[H]
\centering
\caption{UPS output-inverter (grid-forming) parameters}
\label{tab:dc_inv_params}
\begin{tabularx}{\linewidth}{@{}lXr@{}}
\toprule
Symbol & Description & Value \\
\midrule
$f_{\mathrm{ref}}$ & frequency reference & $60$~Hz \\
$V_{\mathrm{AC,ord}}$ & AC voltage order & $0.97$~pu \\
$P_{\mathrm{set}},\ Q_{\mathrm{set}}$ & droop power set points & $0$, $0$~pu \\
$D_{pf}$ & $P$--$f$ droop gain (damping $1000$) & $1\times10^{-3}$ \\
$D_{qV}$ & $Q$--$V$ droop gain (damping $1000$) & $1\times10^{-3}$ \\
$D_{VV}$ & $V_{\mathrm{AC}}$--$V_{\mathrm{DC}}$ droop gain (damping $0.5$) & $2.0$ \\
$k_{P,V}^{\mathrm{ups}}$ & voltage-loop proportional gain (this study) & $1.0$ \\
$k_{I,V}^{\mathrm{ups}}$ & voltage-loop integral gain (this study) & $100$~s$^{-1}$ \\
--- & PWM-level current limit (not in the average model) & $1.5$~pu \\
\bottomrule
\end{tabularx}
\end{table}

\subsection{Smoothing and Discretization}

\begin{table}[H]
\centering
\caption{Smoothing and integration settings}
\label{tab:num_params}
\begin{tabularx}{\linewidth}{@{}lXr@{}}
\toprule
Symbol & Description & Value \\
\midrule
$\varepsilon$ & smoothing parameter of $\smax$, $\smin$, $\sclip$ & $10^{-2}$ \\
$\varepsilon$ & smoothing parameter of $\ssign$, $|\cdot|_\varepsilon$ & $10^{-6}$ \\
$h$ & fast-subsystem step size, $[h^{\min},h^{\max}]$ & $[0.1,1]$~ms \\
$h$ & slow-subsystem step size, $[h^{\min},h^{\max}]$ & $[0.1,10]$~ms \\
\bottomrule
\end{tabularx}
\end{table}

\bibliographystyle{IEEEtran}
\bibliography{biblography_2}

\end{document}